\documentclass[prl,superscriptaddress,showpacs,longbibliography,reprint]{revtex4-2} 
\usepackage{packages}

\newcommand{\id}{\mathds{1}}

\begin{document}
\raggedbottom

\title{Inferring interpretable dynamical generators of local quantum observables from projective measurements through machine learning}

\newcommand{\tubingen}{Institut f\"{u}r Theoretische Physik,  Universit\"{a}t T\"{u}bingen, Auf der Morgenstelle 14, 72076 T\"{u}bingen, Germany}

\author{Giovanni Cemin}
\affiliation{\tubingen}

\author{Francesco Carnazza}
\affiliation{\tubingen}

\author{Sabine Andergassen}
\affiliation{Institute for Solid State Physics and Institute of Information Systems Engineering, Vienna University of Technology, 1040 Vienna, Austria}

\author{Georg Martius}
\affiliation{Max Planck Institute for Intelligent Systems, Max-Planck-Ring 4, 72076 Tübingen, Germany}
\affiliation{Wilhelm Schickard Institut für
Informatik, Maria-von-Linden-Straße 6
72076 T\"{u}bingen}

\author{Federico Carollo}
\affiliation{\tubingen}

\author{Igor Lesanovsky}
\affiliation{\tubingen}
\affiliation{School of Physics and Astronomy and Centre for the Mathematics and Theoretical Physics of Quantum Non-Equilibrium Systems, The University of Nottingham, Nottingham, NG7 2RD, United Kingdom}

\date{\today}

\begin{abstract}
To characterize the dynamical behavior of many-body quantum systems, one is usually interested in the evolution of so-called order-parameters rather than in characterizing the full quantum state. 
In many situations, these quantities coincide with the expectation value of local observables, such as the magnetization or the particle density. 
In experiment, however, these expectation values can only be obtained with a finite degree of accuracy due to the effects of the projection noise. Here, we utilize a machine-learning approach to infer the dynamical generator governing the evolution of local observables in a many-body system from noisy data. To benchmark our method, we consider a variant of the quantum Ising model and generate synthetic experimental data, containing the results of $N$ projective measurements at $M$ sampling points in time, using the time-evolving block-decimation algorithm. As we show, across a wide range of parameters the dynamical generator of local observables can be approximated by a Markovian quantum master equation. Our method is not only useful for extracting effective dynamical generators from many-body systems, but may also be applied for inferring decoherence mechanisms of quantum simulation and computing platforms.
\end{abstract}

\maketitle

\noindent \textbf{Introduction.---}
\begin{figure*}[t!]
    \centering
    \includegraphics[width=\textwidth]{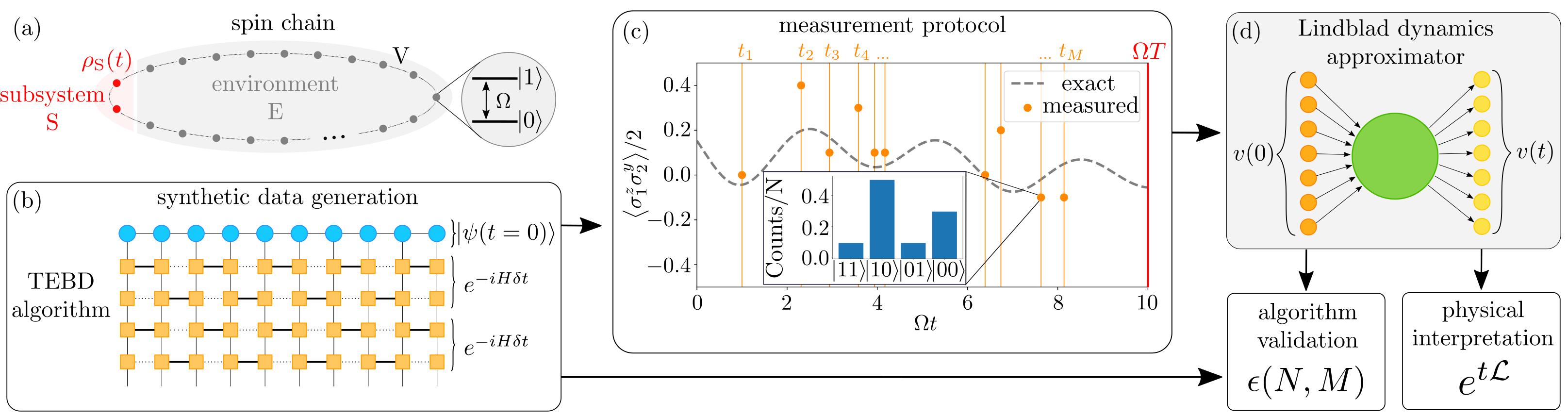}
    \caption{{\bf System and measurement protocol.} (a) The system is composed of the two-spin \textit{subsystem} S, which is embedded in the \textit{environment} E consisting of $L-2$ spins. The Hamiltonian describing the system [cf.~Eq.~\eqref{eq_hamiltonian_spin_chain}] features a driving field with Rabi frequency $\Omega$ and nearest-neighbor interactions with coupling strength $V$.
    (b) Synthetic data generation using the time-evolving block-decimation algorithm. The time evolution operator is approximated using Trotterization: $U_t = e^{-iHt} \approx \prod e^{-iH\delta t}$. 
    (c) Sketch of the measurement protocol. The gray dashed line represents the \textit{exact} dynamics of the observable. For each trajectory, $M$ random points are selected in the time window $[0, \Omega T]$ (here $\Omega T=10$). For each point in time, we measure $N$ times in a suitable basis; here $z$($y$)-basis for the first (second) spin. The inset shows an example histogram of measurement outcomes: $\ket{10}$ means the measurement on the first spin along $z$-basis has outcome $1$, and the measurement on the second spin along $y$-basis has outcome $0$. From the histogram, the expectation value depicted as orange dots are obtained. 
    (d) The synthetic data is fed to the ML algorithm, namely a single layer perceptron, that learns the dynamical generator $\mathcal{L}$. The algorithm is validated by comparison to the exact data, where the error $\epsilon(N,M)$ is calculated for different values of $N$ and $M$. The ML method is interpretable and can be used to  “read  out”  the  underlying  dynamical  processes.
    }
    \label{fig_system}
\end{figure*}
Reconstructing Hamiltonian operators or dynamical generators from physical properties of a quantum system is a problem of current interest. For instance, inverse methods can be applied to identify quantum Hamiltonians associated with a given ground state \cite{Chertkov2018} and interacting many-body theories can be obtained from the knowledge of correlation functions \cite{Zache2020,merger2023}. In many settings, one is merely interested in reconstructing effective equations of motion for a subsystem S embedded in a larger ``environment" E, as it happens, for open quantum systems (OQS) \cite{Liu2011,Navarrete_Benlloch2011,Ma2012,Everest2017,Hoeppe2012}. Furthermore, in the framework of quantum simulation \cite{Georgescu2014,Weimer2010,2008simulating,cirac2012goals,blatt2012quantum,babbush2023quantum,jin2023quantum,zhang2023,Schfer2020}, it is important to understand the (effective) dynamical equations under which artificial quantum systems actually evolve and by how much these differ from the desired ones \cite{quantum_verification,hauke2012}. This is relevant for improving state-of-the-art hardware \cite{PRXQuantum.2.017003,Morgado2021} and the identification of noise models. Another interesting instance concerns the evolution of order parameters, often constructed from local observables, such as the particle density or the magnetization.

Machine learning (ML) approaches appear to be particularly suited for this task  \cite{gebhart2023}. Quantum process tomography with generative adversarial methods \cite{braccia2022}, neural networks \cite{han2021tomography}, and recurrent neural networks \cite{mohseni2022deep,genois2021quantum} have been developed. These approaches are promising but have two main drawbacks: they require a great number of measurements, and they treat ML algorithms as black boxes, thus lacking in physical interpretation. 
Simpler methods are capable of learning Hamiltonians from fewer local measurements \cite{Wilde2022, Bairey2019,di2009hamiltonian,cole2005identifying,devitt2006scheme,wang2017experimental,gentile2021learning}, yet they typically rely on a \textit{a priori} ansatz for the functional form of the Hamiltonian or of the dissipation.
A more general approach is to fit an OQS dynamics by learning the Nakajima-Zwanzig equation \cite{nakajima1958,zwanzig1960} through transfer tensor techniques \cite{cerrillo2014,gelzinis2017,pollock2018} or by learning convolutionless master equations \cite{banchi2018modelling}. However, these approaches require a full state tomography at different time steps, which is prohibitive to achieve in experiments.
Ultimately, current methods thus either rely on an \textit{ad hoc} ansatz, or demand data which is not experimentally accessible, or lack in physical interpretability (which is actually becoming highly desirable \cite{link2023}).

In this work, we use ML methods to infer the effective dynamical generator of a subsystem from a finite set of local measurements at randomly selected times, which inevitably produce noisy data due to projection noise. To illustrate our approach, we consider a many-body spin system [cf.~Fig.~\ref{fig_system}(a)], which is ubiquitous in the context of experiments with trapped ion or Rydberg atom quantum simulators \cite{Morgado2021,Wu_2021,Saffman2010}. By using synthetic (experimental) data generated with tensor-network based algorithms we infer a physically consistent Markovian dynamical generator \cite{Mazza2021,Carnazza2022} governing the evolution of a small subsystem. Our method, which works reliably across a wide range of parameters -- even in some instances outside the weak coupling limit -- yields interpretable results which may be used to infer noise models on quantum simulators or to study thermalization dynamics in many-body systems. 
\\

\noindent \textbf{Setting.---}
The system we consider is a $1D$ quantum spin chain consisting of $L$ spins arranged on a circular lattice, as depicted in Fig.~\ref{fig_system}(a). The chain is partitioned into a \textit{subsystem} S, here formed by two adjacent spins, and the \textit{environment} E, that is, the remainder of the spin chain. We assume the whole system to evolve unitarily, through the many-body Hamiltonian
\begin{equation}
    H_{S+E} = \frac{\Omega}{2} \sum_{i=1}^{L} \sigma_i^x + V\bigg( \sum_{i=1}^{L-1} n_i n_{i+1} + n_L n_{1}\bigg)\, .
    \label{eq_hamiltonian_spin_chain}
\end{equation}
The first term describes a transverse ``laser" field, while the second one accounts for nearest-neighbor interactions. Here, $\sigma_i^\alpha$ denotes the $\alpha$ Pauli matrix for the $i^{th}$ spin and we have defined the projector $n = \frac{1+\sigma^z}{2}$. The above Hamiltonian is of practical interest for experiments with Rydberg atoms \cite{Saffman2010} and essentially encodes an Ising model in the presence of transverse and longitudinal fields. We simulate the time evolution of the whole system by means of the time-evolving block-decimation (TEBD) algorithm [see Fig.~\ref{fig_system}(b)]. 

In our setting, the information on the state of S is obtained by a finite number, $N$, of projective measurements, taken at randomly selected times $t_1,... \,t_M$ [see Fig.~\ref{fig_system}(c)]. From this \textit{noisy} data we want to infer the open quantum dynamics of the reduced state $\rho_{\mathrm{S}}(t)$ of subsystem S. Formally, this dynamics is obtained as the partial trace of the evolution of the full many-body state, i.e., $\rho_{\mathrm{S}}(t)={\rm Tr}_{\mathrm{E}}\left(U_t \rho_{\mathrm{S+E}}U_t^\dagger\right)$, where $U_t=e^{-iHt}$, $\rho_{\mathrm{S+E}}$ is the initial state of the system and ${\rm Tr}_{\mathrm{E}}$ denotes the trace over the environment. In general, such a dynamics is rather involved and may show non-Markovian effects or it may be nonlinear for generic initial states $\rho_{\mathrm{S+E}}$ \cite{BRE02}. Here, we restrict ourselves to learning a Markovian dynamics for $\rho_{\mathrm{S}}(t)$, but more general approaches are possible \cite{Mazza2021}. The goal is then to identify the time-independent generator $\mathcal{L}$, yielding the Markovian quantum master equation evolution \cite{BRE02,Gorini1976,Lindblad1976},
\begin{equation}
    \dot{\rho}_S(t) = \mathcal{L}[\rho_S(t)] 
    \label{eq_master_eq_formal}
\end{equation}
that {\it optimally} describes the dynamics of S. This simple form has the advantage that it is interpretable, i.e., it allows to read off the Hamiltonian and decoherence processes (see further below).\\

\noindent \textbf{Data generation.---} We simulate the time evolution of the system for times $t\in[0, \Omega T]$ by means of the TEBD algorithm \cite{Vidal2004,Zwolak2004,gray2018quimb} [see Fig.~\ref{fig_system}(b)], which allows us to study systems of up to $50$ spins. 
We generate $30$ trajectories obtained by initializing the system in state $\psi=\bigotimes_{k=1}^L \ket{0}$, with $\sigma^z\ket{0}=-\ket{0}$ and perturbing the subsystem S through a random two-spin unitary $\hat{U}_{\mathrm{rand}}$ distributed with the Haar measure. 
As the system evolves in time, at intervals of $\Omega dt=0.01$, we compute the expectation value of the $15$ independent observables $\{ \id_1 \otimes \sigma_2^x, \id_1 \otimes \sigma_2^y, ..., \sigma_1^x \otimes \sigma_2^y, ..., \sigma_1^z \otimes \sigma_2^z \}/2$ of the subsystem S, which uniquely identify the reduced subsystem state. We then emulate experimental measurements of these local observables. For each trajectory, we select $M$ random times in the time-window $[0,\Omega T]$ and, for each of these points in time, we perform $N$ measurements in all the relevant basis in order to produce a noisy estimate of the reduced state.
Such a procedure is sketched in Fig.~\ref{fig_system}(c), where the histogram depicts the counting of $N=10$ measurement outcomes for a single time-point, whereas the orange dots represent the experimental expectation values.\\

\noindent \textbf{ML architecture and training.---}
The generator $\mathcal{L}$ defining the quantum master equation \eqref{eq_master_eq_formal} can be parametrized as $\mathcal{L}=\mathcal{H}+\mathcal{D}$, where
\begin{equation}
    \begin{aligned}
        \mathcal{H}[\cdot] &= -i [H,\cdot]\, , \quad \mbox{with} \quad H= \sum_{i=2}^{d^2} \theta^H_i F_i \, ,\\
        \mathcal{D}[\cdot] &= \frac{1}{2} \sum_{i,j=2}^{d^2} c_{ij} ([F_i, \cdot F^{\dagger}_j] + [F_i \cdot, F^{\dagger}_j]) \, .
    \end{aligned}
    \label{eq_Lindblad_explicit}
\end{equation}
Here, we have introduced the Hermitian orthonormal basis $\{F_i\}_{i=1}^{16}$, for the operators of the subsystem S, where we choose $F_1$ to be proportional to the identity operator, $F_1 = \id/\sqrt{d}$. Moreover, we write the Hamiltonian in terms of the ``fields" $\theta^H_i$, and the dissipative contribution $\mathcal{D}$ in a non-diagonal form, fully specified by the so-called Kossakowski matrix $c_{ij}$.  The latter must be positive semi-definite in order for the open quantum dynamics to be completely positive. This constraint can be ``hard coded" by setting $c = Z^{\dagger} Z$, for a complex matrix $Z = \theta^{X}+i\theta^{Y}$, with $\theta^{X}$ and $\theta^{Y}$ being real-valued.

We decompose the reduced state $\rho_S$ on the basis $\{F_i\}_{i=1}^{16}$ as \cite{Byrd2003}
\begin{equation}
    \rho_{\mathrm{S}} = \frac{\id}{d} + \sum_{i=2}^{16} F_i v_i \, ,
    \label{eq_rho_decomposition}
\end{equation}
which defines the {\it coherence vector} $v_i = \Tr(F_i \rho_{\mathrm{S}})$. Notice the condition $\Tr(\rho_{\mathrm{S}})=1$ implies $v_1 = \frac{1}{\sqrt{d}}$, that we take outside the sum. 
The coherence vector representation is quite convenient, from a numerical point of view, as it allows us to write the action of the generator on states as the action of the matrix $\mathbf{L}$ on coherence vectors. The quantum master equation \eqref{eq_master_eq_formal} becomes
\begin{equation}
    \frac{d \mathbf{v}(t)}{d t} = \mathbf{L} \mathbf{v}(t) = ( \mathbf{H} + \mathbf{D} ) \mathbf{v}(t) \, ,
    \label{eq_master_eq}
\end{equation}
where $\mathbf{H}_{ij}=-\Tr(\mathcal{H}[F_i]F_j)$ and $\mathbf{D}_{ij}=\Tr(\mathcal{D}[F_i]F_j)$ are real-valued matrices. 
As explicitly shown in the Supplemental Material (see Ref.~\cite{SM}), the matrix $\mathbf{H}=\mathbf{H}(\theta^{H}_i)$ depends linearly on the parameters $\theta^{H}_i$, while the matrix $\mathbf{D}=\mathbf{D}(\theta^X_{ij},\theta^Y_{ij})$ depends quadratically on $\theta^X_{ij}$ and $\theta^Y_{ij}$. 

We build a simple neural network \cite{pytorch}, here called {\it Lindblad dynamics approximator} (LDA), as
\begin{equation}
    \mathbf{M}(\theta, t) = e^{t [ \mathbf{H}(\theta^{H}_i) + \mathbf{D}(\theta^X_{ij},\theta^Y_{ij}) ] } \, ,
\end{equation}
that is the structure of the Lindblad time propagator.

We train the LDA to learn the Lindblad representation $\mathbf{L}$ from (synthetic) experimental data. In the training procedure, we feed the LDA with the initial conditions $\mathbf{v}_{\mathrm{in}} = \mathbf{v}(0)$ and the time of the measurement $t$, and optimize the parameters $\theta=\{\theta^H_i,\theta^X_{ij},\theta^Y_{ij}\}$ such that $\mathbf{v}_{\mathrm{out}} \simeq \mathbf{v}(t)=\mathbf{M}(\theta,t)\mathbf{v}_{\mathrm{in}}$ \cite{adam}. 
Training over a finite time $t$ is crucial when working with  experimental data. Indeed, training the LDA to propagate the coherence vector only over an infinitesimal time-step $dt$ \cite{Carnazza2022,Mazza2021}, i.e.,  $\mathbf{v}_{\mathrm{in}} = \mathbf{v}(t)$ and $\mathbf{v}_{\mathrm{out}} = \mathbf{v}(t+dt)$, is bound to fail as soon as the noise is larger than the variation of the coherence vector. For further details see Supplemental Material \cite{SM}.

To test the correctness of the learned generator, we produce $r$ new exact trajectories and compare them with the LDA prediction.
A quantitative measure of the performance of the ML algorithm is given by the following error function
\begin{equation}
    \epsilon(N,M) := \frac{1}{r} \sum^{r}_{i=1} \frac{1}{T} \int^{T}_{0} \frac{\norm{\rho^i_{\mathrm{ML}}(t) - \rho^i_{\mathrm{S}}(t)}_2^2}{\norm{\rho^i_{\mathrm{S}}(t)}_2^2} \dd t \, ,
    \label{eq_error_model}
\end{equation}
where $\rho^i_{\mathrm{ML}}(t)$ is the prediction for the state of the subsystem obtained from our ML algorithm, $\rho^i_{\mathrm{S}}(t)$ represents the synthetic data for a given choice of $N$ and $M$, and $\norm{O}_2^2={\rm Tr}\left(O^\dagger O\right)$\footnote{The code for the generation of the artificial data and the training of the ML algorithm is made available at \href{https://github.com/giovannicemin/lindblad_dynamics_approximator}{github.com/giovannicemin/lindblad\_dynamics\_approximator}.}.\\

\noindent \textbf{Benchmarking the algorithm.---}
\begin{figure}[t]
    \centering
    \includegraphics[width=\columnwidth]{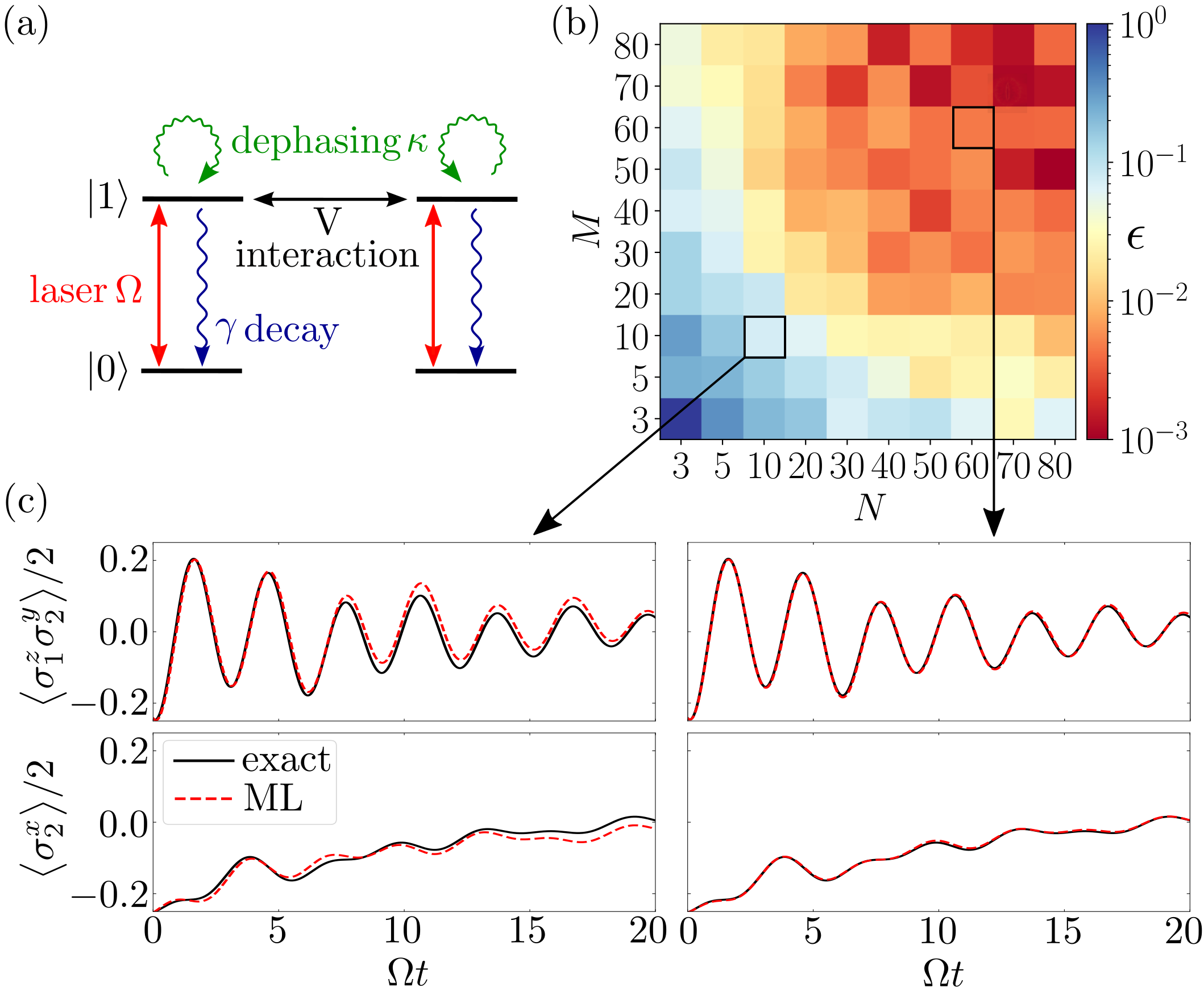}
    \caption{{\bf Benchmark.} (a) OQS used to benchmark the LDA, described in terms of a Hamiltonian and jump operators [cf.~Eq.~\eqref{benchmark_model}]. The Hamiltonian consists of a driving ($\Omega$) and an interaction ($V$). Jump operators describe decay (with rate $\gamma$) and dephasing (with rate $\kappa$).
    (b) The error $\epsilon (N,M)$ as in Eq.~\eqref{eq_error_model} plotted for different values of $N$ and $M$. The error is averaged over $r=10$ out-of-sample trajectories with a doubled time-window $[0, 2 \Omega T]$ (see main text for details). 
    (c) Dynamical curves for two selected observables and different combinations of $N$ and $M$. From the plot one can appreciate how the predicted dynamics is already quite reliable for smaller values of $N$ and $M$. The parameters are $V=0.5\Omega$, $\gamma = 0.01\Omega$ and $\kappa=0.05\Omega$. }
    \label{fig_benchmark}
\end{figure}
Before training the algorithm on data for the many-body model in Eq.~\eqref{eq_hamiltonian_spin_chain}, we benchmark its ability to learn a Lindblad generator within a well-controlled setting.
We consider a two-spin Lindblad generator, which in its diagonal form, is specified by the following Hamiltonian and jump operators [cf.~Fig.~\ref{fig_benchmark}(a)]
\begin{align}
    H &= \frac{\Omega}{2} (\sigma_1^x + \sigma_2^x ) + V n_1 n_2 \, , \nonumber\\
    J_1 &= \sqrt{\gamma} \sigma_1^- \, , \qquad J_2 = \sqrt{\gamma} \sigma_2^- \, , \label{benchmark_model}\\
    J_3 &= \sqrt{\kappa} n_1 \, , \qquad J_4 = \sqrt{\kappa} n_2\, . \nonumber
\end{align}
The jump operators effectively describe the effects of the environment on the subsystem: $J_1, J_2$ encode decay from $\ket{1}$ to $\ket{0}$ while $J_3,J_4$ encode dephasing.
The data generation procedure follows the protocol described above, with two exceptions: the initial conditions are given by a random density matrix $\rho_{\mathrm{rand}}$, and the data for testing have a doubled time-window $[ 0, 2\Omega T ]$. In this way, we can also test the ability of the algorithm to extrapolate to unseen times. Since the network is, in principle, able to perfectly learn a Markovian Lindblad generator, we expect the extrapolation to be accurate.  
The results are reported in Fig.~\ref{fig_benchmark} (b-c). The color map [panel (b)] shows the error, $\epsilon(N,M)$, averaged over $r=10$ test trajectories, for 100 different combinations of $N$ and $M$. 
First, we observe the overall trend of the error to decrease as $N\times M$ increases. However, the plot is not symmetric with respect to the line  $M=N$. In fact, for higher values of $M$ (and fixed $N\times M$), the LDA has a more stable training, and hence a better performance. This is because the choice of the $M$ time-points is random, and a small value of $M$ has a high probability of yielding data which is not representative of the dynamics. On the other hand, higher values of $M$ yields more representative data of the trajectory, despite relatively small values of $N$. 
For high values of $N$ and $M$, as expected, the error becomes small. 
The ML algorithm can thus successfully learn a Lindbladian, with a precision that approximately depends on the product $N\times M$. \\

\noindent \textbf{Many-body setting.---}
\begin{figure}[t!]
    \centering
    \includegraphics[width=\columnwidth]{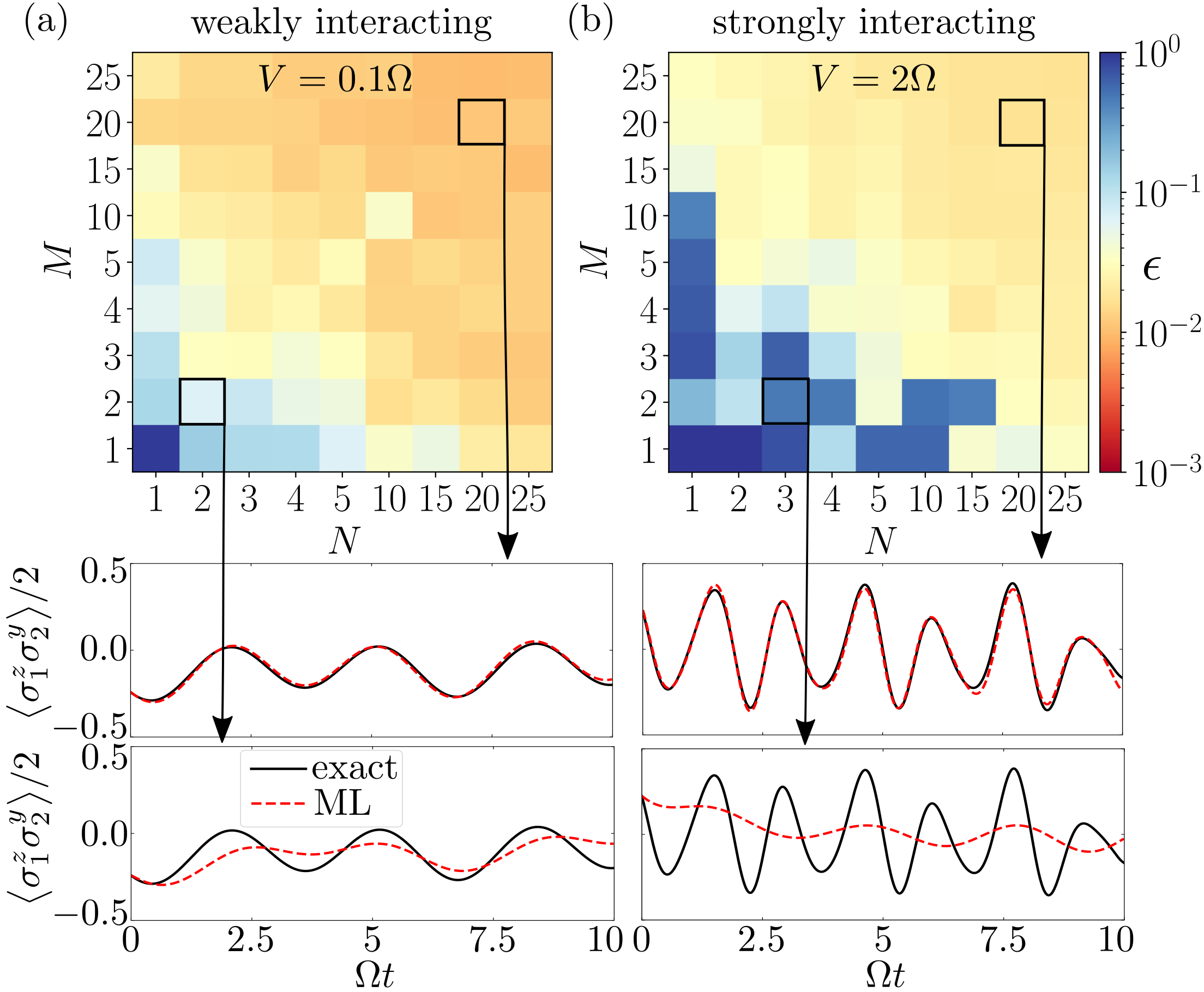}
    \caption{{\bf Learning the subsystem dynamics.} Error $\epsilon(N,M)$ as in Eq.~\eqref{eq_error_model} for  different values of $N$ and $M$, for two different combinations of the parameters. The error is calculated over $r=10$ trajectories with $t \in [0, \Omega T]$. (a) Data for the case $V=0.1\Omega$. Here, the dynamics is slow and the ML algorithm learns an effective $\mathcal{L}$ already for small values of $N$ and $M$. (b) Data for $V=2\Omega$. In this case, two different regimes can be distinguished neatly: either the model does not learn the dynamics (blue squares) or it learns it (yellow squares).
    In both cases, the errors are due to the non-Markovianity assumed for the learned model.
    The plots beneath the color maps reports the time evolution of an observable for the two different models and different sets of $(N,M)$.}
    \label{fig_results}
\end{figure}
Having established the capability of the LDA to {\it exactly} learn Lindblad generators from experimental data, we now address the many-body scenario described by Eq.~\eqref{eq_hamiltonian_spin_chain}. In this case, the reduced dynamics of $\rho_{\rm S}(t)$ can feature non-Markovian effects. The LDA will thus, by construction, learn the ``optimal" Markovian description of the system. Whether this  description will accurately describe the subsystem or not depends on the relevance of non-Markovian effects. 

In Fig.~\ref{fig_results}(a), we show results for weak interactions $V=0.1\Omega$. Here, the ML algorithm learns a dynamical generator which faithfully reproduces the time evolution of the coherence vector. This suggests that the subsystem dynamics is, in this weak-coupling regime, essentially Markovian.  Notably, the Lindblad generator can be inferred even when $N$ and $M$ are small.
As already observed during the benchmarking, the training procedure is less stable for the models in the bottom-right corner of Fig.~\ref{fig_results}(a), compared to the top left, confirming that larger $M$ values are better than larger $N$ values at fixed $N\times M$.

In Fig.~\ref{fig_results}(b) we show results for strong interactions, $V=2\Omega$. Quite surprisingly, also in this case, the LDA learns an effective Lindbladian which reproduces very well the subsystem dynamics. Also here, the latter has a Markovian character. A possible explanation for this is that strong interactions lead to a faster decay of time correlations in the environment, which thus renders the subsystem dynamics Markovian.
Due to the faster oscillations in this regime, more sampling points in time are needed than the weakly interacting case, Fig.~\ref{fig_results}(a), for a same accuracy. 
In both cases, for small $N \times M$, while the model cannot recover the exact dynamics, it nonetheless provides an average over the fast oscillations  [cf.~Figs.~\ref{fig_results}(a)-\ref{fig_results}(b)]. The data for the whole coherence vector are reported in the supplemental material \cite{SM}, where we also show additional results on the case $V=0.5\Omega$. In the latter case, the error is higher, due to non-Markovian effects which appear to be non-negligible \cite{Roos2020} in this intermediate regime.
\\

\noindent \textbf{Conclusions.---}
We have presented a simple ML algorithm able to learn a physically consistent and interpretable dynamical generator starting from (synthetic) experimental data. We have shown that it can yield faithful results for both weak and strong interactions. A wider range of systems could be included by relaxing the assumed Markovianity of the learned generator, namely by allowing it  to be time-dependent $\mathcal{L}(t)$. Some approaches already exist (see, e.g., Refs.~\cite{banchi2018modelling,cerrillo2014}), but they require an enormous amount of data and lack physical interpretability.

In this work, we restricted the study to 1D systems with NN interactions. However, our ML method yields physically consistent generators regardless of the dimensionality and the range of the interactions, to which the model is completely agnostic. Moreover, it is interpretable, hence it can be used to ``read out" the underlying dynamical processes (see Supplemental Material \cite{SM}). In fact, the learned matrix $\mathbf{L}$ gives direct access to the parameters $\theta^H_i,\theta^X_{ij},\theta^Y_{ij}$, which represent the Hamiltonian and the jump operators of the subsystem S.
In the future, it would be interesting to understand whether feeding the ML algorithm with a sampling of the full coherence vector is necessary or whether a \textit{bona fide} dynamics can still be learned when leaving out information about certain observables. \\ 

\textbf{Acknowledgments.---} We acknowledge financial support from the Deutsche Forschungsgemeinschaft (DFG, German Research Foundation) under Germany's Excellence Strategy—EXC-Number 2064/1-Project Number 390727645, under Project No. 449905436 and through the Research Unit FOR 5413/1, Grant No. 465199066. This project has also received funding from the European Union’s Horizon Europe research and innovation program under Grant Agreement No. 101046968 (BRISQ). FC~is indebted to the Baden-W\"urttemberg Stiftung for the financial support of this research project by the Eliteprogramme for Postdocs.

\bibliography{refs}

\begin{thebibliography}{56}%
\makeatletter
\providecommand \@ifxundefined [1]{%
 \@ifx{#1\undefined}
}%
\providecommand \@ifnum [1]{%
 \ifnum #1\expandafter \@firstoftwo
 \else \expandafter \@secondoftwo
 \fi
}%
\providecommand \@ifx [1]{%
 \ifx #1\expandafter \@firstoftwo
 \else \expandafter \@secondoftwo
 \fi
}%
\providecommand \natexlab [1]{#1}%
\providecommand \enquote  [1]{``#1''}%
\providecommand \bibnamefont  [1]{#1}%
\providecommand \bibfnamefont [1]{#1}%
\providecommand \citenamefont [1]{#1}%
\providecommand \href@noop [0]{\@secondoftwo}%
\providecommand \href [0]{\begingroup \@sanitize@url \@href}%
\providecommand \@href[1]{\@@startlink{#1}\@@href}%
\providecommand \@@href[1]{\endgroup#1\@@endlink}%
\providecommand \@sanitize@url [0]{\catcode `\\12\catcode `\$12\catcode
  `\&12\catcode `\#12\catcode `\^12\catcode `\_12\catcode `\%12\relax}%
\providecommand \@@startlink[1]{}%
\providecommand \@@endlink[0]{}%
\providecommand \url  [0]{\begingroup\@sanitize@url \@url }%
\providecommand \@url [1]{\endgroup\@href {#1}{\urlprefix }}%
\providecommand \urlprefix  [0]{URL }%
\providecommand \Eprint [0]{\href }%
\providecommand \doibase [0]{https://doi.org/}%
\providecommand \selectlanguage [0]{\@gobble}%
\providecommand \bibinfo  [0]{\@secondoftwo}%
\providecommand \bibfield  [0]{\@secondoftwo}%
\providecommand \translation [1]{[#1]}%
\providecommand \BibitemOpen [0]{}%
\providecommand \bibitemStop [0]{}%
\providecommand \bibitemNoStop [0]{.\EOS\space}%
\providecommand \EOS [0]{\spacefactor3000\relax}%
\providecommand \BibitemShut  [1]{\csname bibitem#1\endcsname}%
\let\auto@bib@innerbib\@empty
\bibitem [{\citenamefont {Chertkov}\ and\ \citenamefont
  {Clark}(2018)}]{Chertkov2018}%
  \BibitemOpen
  \bibfield  {author} {\bibinfo {author} {\bibfnamefont {E.}~\bibnamefont
  {Chertkov}}\ and\ \bibinfo {author} {\bibfnamefont {B.~K.}\ \bibnamefont
  {Clark}},\ }\bibfield  {title} {\bibinfo {title} {Computational inverse
  method for constructing spaces of quantum models from wave functions},\
  }\href {https://doi.org/10.1103/PhysRevX.8.031029} {\bibfield  {journal}
  {\bibinfo  {journal} {Phys. Rev. X}\ }\textbf {\bibinfo {volume} {8}},\
  \bibinfo {pages} {031029} (\bibinfo {year} {2018})}\BibitemShut {NoStop}%
\bibitem [{\citenamefont {Zache}\ \emph {et~al.}(2020)\citenamefont {Zache},
  \citenamefont {Schweigler}, \citenamefont {Erne}, \citenamefont
  {Schmiedmayer},\ and\ \citenamefont {Berges}}]{Zache2020}%
  \BibitemOpen
  \bibfield  {author} {\bibinfo {author} {\bibfnamefont {T.~V.}\ \bibnamefont
  {Zache}}, \bibinfo {author} {\bibfnamefont {T.}~\bibnamefont {Schweigler}},
  \bibinfo {author} {\bibfnamefont {S.}~\bibnamefont {Erne}}, \bibinfo {author}
  {\bibfnamefont {J.}~\bibnamefont {Schmiedmayer}},\ and\ \bibinfo {author}
  {\bibfnamefont {J.}~\bibnamefont {Berges}},\ }\bibfield  {title} {\bibinfo
  {title} {Extracting the field theory description of a quantum many-body
  system from experimental data},\ }\href
  {https://doi.org/10.1103/PhysRevX.10.011020} {\bibfield  {journal} {\bibinfo
  {journal} {Phys. Rev. X}\ }\textbf {\bibinfo {volume} {10}},\ \bibinfo
  {pages} {011020} (\bibinfo {year} {2020})}\BibitemShut {NoStop}%
\bibitem [{\citenamefont {Merger}\ \emph {et~al.}(2023)\citenamefont {Merger},
  \citenamefont {Ren{\'e}}, \citenamefont {Fischer}, \citenamefont {Bouss},
  \citenamefont {Nestler}, \citenamefont {Dahmen}, \citenamefont {Honerkamp},\
  and\ \citenamefont {Helias}}]{merger2023}%
  \BibitemOpen
  \bibfield  {author} {\bibinfo {author} {\bibfnamefont {C.}~\bibnamefont
  {Merger}}, \bibinfo {author} {\bibfnamefont {A.}~\bibnamefont {Ren{\'e}}},
  \bibinfo {author} {\bibfnamefont {K.}~\bibnamefont {Fischer}}, \bibinfo
  {author} {\bibfnamefont {P.}~\bibnamefont {Bouss}}, \bibinfo {author}
  {\bibfnamefont {S.}~\bibnamefont {Nestler}}, \bibinfo {author} {\bibfnamefont
  {D.}~\bibnamefont {Dahmen}}, \bibinfo {author} {\bibfnamefont
  {C.}~\bibnamefont {Honerkamp}},\ and\ \bibinfo {author} {\bibfnamefont
  {M.}~\bibnamefont {Helias}},\ }\href@noop {} {\bibinfo {title} {Learning
  interacting theories from data}} (\bibinfo {year} {2023}),\ \Eprint
  {https://arxiv.org/abs/2304.00599} {arXiv:2304.00599} \BibitemShut {NoStop}%
\bibitem [{\citenamefont {Liu}\ \emph {et~al.}(2011)\citenamefont {Liu},
  \citenamefont {Li}, \citenamefont {Huang}, \citenamefont {Li}, \citenamefont
  {Guo}, \citenamefont {Laine}, \citenamefont {Breuer},\ and\ \citenamefont
  {Piilo}}]{Liu2011}%
  \BibitemOpen
  \bibfield  {author} {\bibinfo {author} {\bibfnamefont {B.-H.}\ \bibnamefont
  {Liu}}, \bibinfo {author} {\bibfnamefont {L.}~\bibnamefont {Li}}, \bibinfo
  {author} {\bibfnamefont {Y.-F.}\ \bibnamefont {Huang}}, \bibinfo {author}
  {\bibfnamefont {C.-F.}\ \bibnamefont {Li}}, \bibinfo {author} {\bibfnamefont
  {G.-C.}\ \bibnamefont {Guo}}, \bibinfo {author} {\bibfnamefont {E.-M.}\
  \bibnamefont {Laine}}, \bibinfo {author} {\bibfnamefont {H.-P.}\ \bibnamefont
  {Breuer}},\ and\ \bibinfo {author} {\bibfnamefont {J.}~\bibnamefont
  {Piilo}},\ }\bibfield  {title} {\bibinfo {title} {Experimental control of the
  transition from {Markovian} to non-{Markovian} dynamics of open quantum
  systems},\ }\href {https://doi.org/10.1038/nphys2085} {\bibfield  {journal}
  {\bibinfo  {journal} {Nat. Phys.}\ }\textbf {\bibinfo {volume} {7}},\
  \bibinfo {pages} {931} (\bibinfo {year} {2011})}\BibitemShut {NoStop}%
\bibitem [{\citenamefont {Navarrete-Benlloch}\ \emph
  {et~al.}(2011)\citenamefont {Navarrete-Benlloch}, \citenamefont {de~Vega},
  \citenamefont {Porras},\ and\ \citenamefont
  {Cirac}}]{Navarrete_Benlloch2011}%
  \BibitemOpen
  \bibfield  {author} {\bibinfo {author} {\bibfnamefont {C.}~\bibnamefont
  {Navarrete-Benlloch}}, \bibinfo {author} {\bibfnamefont {I.}~\bibnamefont
  {de~Vega}}, \bibinfo {author} {\bibfnamefont {D.}~\bibnamefont {Porras}},\
  and\ \bibinfo {author} {\bibfnamefont {J.~I.}\ \bibnamefont {Cirac}},\
  }\bibfield  {title} {\bibinfo {title} {Simulating quantum-optical phenomena
  with cold atoms in optical lattices},\ }\href
  {https://doi.org/10.1088/1367-2630/13/2/023024} {\bibfield  {journal}
  {\bibinfo  {journal} {New J. Phys.}\ }\textbf {\bibinfo {volume} {13}},\
  \bibinfo {pages} {023024} (\bibinfo {year} {2011})}\BibitemShut {NoStop}%
\bibitem [{\citenamefont {Ma}\ \emph {et~al.}(2012)\citenamefont {Ma},
  \citenamefont {Sun}, \citenamefont {Wang},\ and\ \citenamefont
  {Nori}}]{Ma2012}%
  \BibitemOpen
  \bibfield  {author} {\bibinfo {author} {\bibfnamefont {J.}~\bibnamefont
  {Ma}}, \bibinfo {author} {\bibfnamefont {Z.}~\bibnamefont {Sun}}, \bibinfo
  {author} {\bibfnamefont {X.}~\bibnamefont {Wang}},\ and\ \bibinfo {author}
  {\bibfnamefont {F.}~\bibnamefont {Nori}},\ }\bibfield  {title} {\bibinfo
  {title} {Entanglement dynamics of two qubits in a common bath},\ }\href
  {https://doi.org/10.1103/PhysRevA.85.062323} {\bibfield  {journal} {\bibinfo
  {journal} {Phys. Rev. A}\ }\textbf {\bibinfo {volume} {85}},\ \bibinfo
  {pages} {062323} (\bibinfo {year} {2012})}\BibitemShut {NoStop}%
\bibitem [{\citenamefont {Everest}\ \emph {et~al.}(2017)\citenamefont
  {Everest}, \citenamefont {Lesanovsky}, \citenamefont {Garrahan},\ and\
  \citenamefont {Levi}}]{Everest2017}%
  \BibitemOpen
  \bibfield  {author} {\bibinfo {author} {\bibfnamefont {B.}~\bibnamefont
  {Everest}}, \bibinfo {author} {\bibfnamefont {I.}~\bibnamefont {Lesanovsky}},
  \bibinfo {author} {\bibfnamefont {J.~P.}\ \bibnamefont {Garrahan}},\ and\
  \bibinfo {author} {\bibfnamefont {E.}~\bibnamefont {Levi}},\ }\bibfield
  {title} {\bibinfo {title} {Role of interactions in a dissipative many-body
  localized system},\ }\href {https://doi.org/10.1103/PhysRevB.95.024310}
  {\bibfield  {journal} {\bibinfo  {journal} {Phys. Rev. B}\ }\textbf {\bibinfo
  {volume} {95}},\ \bibinfo {pages} {024310} (\bibinfo {year}
  {2017})}\BibitemShut {NoStop}%
\bibitem [{\citenamefont {Hoeppe}\ \emph {et~al.}(2012)\citenamefont {Hoeppe},
  \citenamefont {Wolff}, \citenamefont {K{\"u}chenmeister}, \citenamefont
  {Niegemann}, \citenamefont {Drescher}, \citenamefont {Benner},\ and\
  \citenamefont {Busch}}]{Hoeppe2012}%
  \BibitemOpen
  \bibfield  {author} {\bibinfo {author} {\bibfnamefont {U.}~\bibnamefont
  {Hoeppe}}, \bibinfo {author} {\bibfnamefont {C.}~\bibnamefont {Wolff}},
  \bibinfo {author} {\bibfnamefont {J.}~\bibnamefont {K{\"u}chenmeister}},
  \bibinfo {author} {\bibfnamefont {J.}~\bibnamefont {Niegemann}}, \bibinfo
  {author} {\bibfnamefont {M.}~\bibnamefont {Drescher}}, \bibinfo {author}
  {\bibfnamefont {H.}~\bibnamefont {Benner}},\ and\ \bibinfo {author}
  {\bibfnamefont {K.}~\bibnamefont {Busch}},\ }\bibfield  {title} {\bibinfo
  {title} {Direct observation of non-{Markovian} radiation dynamics in {3D}
  bulk photonic crystals},\ }\href
  {https://doi.org/10.1103/PhysRevLett.108.043603} {\bibfield  {journal}
  {\bibinfo  {journal} {Phys. Rev. Lett.}\ }\textbf {\bibinfo {volume} {108}},\
  \bibinfo {pages} {043603} (\bibinfo {year} {2012})}\BibitemShut {NoStop}%
\bibitem [{\citenamefont {Georgescu}\ \emph {et~al.}(2014)\citenamefont
  {Georgescu}, \citenamefont {Ashhab},\ and\ \citenamefont
  {Nori}}]{Georgescu2014}%
  \BibitemOpen
  \bibfield  {author} {\bibinfo {author} {\bibfnamefont {I.~M.}\ \bibnamefont
  {Georgescu}}, \bibinfo {author} {\bibfnamefont {S.}~\bibnamefont {Ashhab}},\
  and\ \bibinfo {author} {\bibfnamefont {F.}~\bibnamefont {Nori}},\ }\bibfield
  {title} {\bibinfo {title} {Quantum simulation},\ }\href
  {https://doi.org/10.1103/RevModPhys.86.153} {\bibfield  {journal} {\bibinfo
  {journal} {Rev. Mod. Phys.}\ }\textbf {\bibinfo {volume} {86}},\ \bibinfo
  {pages} {153} (\bibinfo {year} {2014})}\BibitemShut {NoStop}%
\bibitem [{\citenamefont {Weimer}\ \emph {et~al.}(2010)\citenamefont {Weimer},
  \citenamefont {M{\"u}ller}, \citenamefont {Lesanovsky}, \citenamefont
  {Zoller},\ and\ \citenamefont {B{\"u}chler}}]{Weimer2010}%
  \BibitemOpen
  \bibfield  {author} {\bibinfo {author} {\bibfnamefont {H.}~\bibnamefont
  {Weimer}}, \bibinfo {author} {\bibfnamefont {M.}~\bibnamefont {M{\"u}ller}},
  \bibinfo {author} {\bibfnamefont {I.}~\bibnamefont {Lesanovsky}}, \bibinfo
  {author} {\bibfnamefont {P.}~\bibnamefont {Zoller}},\ and\ \bibinfo {author}
  {\bibfnamefont {H.~P.}\ \bibnamefont {B{\"u}chler}},\ }\bibfield  {title}
  {\bibinfo {title} {A {Rydberg} quantum simulator},\ }\href
  {https://doi.org/10.1038/nphys1614} {\bibfield  {journal} {\bibinfo
  {journal} {Nat. Phys.}\ }\textbf {\bibinfo {volume} {6}},\ \bibinfo {pages}
  {382} (\bibinfo {year} {2010})}\BibitemShut {NoStop}%
\bibitem [{\citenamefont {Friedenauer}\ \emph {et~al.}(2008)\citenamefont
  {Friedenauer}, \citenamefont {Schmitz}, \citenamefont {Glueckert},
  \citenamefont {Porras},\ and\ \citenamefont {Sch{\"a}tz}}]{2008simulating}%
  \BibitemOpen
  \bibfield  {author} {\bibinfo {author} {\bibfnamefont {A.}~\bibnamefont
  {Friedenauer}}, \bibinfo {author} {\bibfnamefont {H.}~\bibnamefont
  {Schmitz}}, \bibinfo {author} {\bibfnamefont {J.~T.}\ \bibnamefont
  {Glueckert}}, \bibinfo {author} {\bibfnamefont {D.}~\bibnamefont {Porras}},\
  and\ \bibinfo {author} {\bibfnamefont {T.}~\bibnamefont {Sch{\"a}tz}},\
  }\bibfield  {title} {\bibinfo {title} {Simulating a quantum magnet with
  trapped ions},\ }\href {https://doi.org/10.1038/nphys1032} {\bibfield
  {journal} {\bibinfo  {journal} {Nat. Phys.}\ }\textbf {\bibinfo {volume}
  {4}},\ \bibinfo {pages} {757} (\bibinfo {year} {2008})}\BibitemShut {NoStop}%
\bibitem [{\citenamefont {Cirac}\ and\ \citenamefont
  {Zoller}(2012)}]{cirac2012goals}%
  \BibitemOpen
  \bibfield  {author} {\bibinfo {author} {\bibfnamefont {J.~I.}\ \bibnamefont
  {Cirac}}\ and\ \bibinfo {author} {\bibfnamefont {P.}~\bibnamefont {Zoller}},\
  }\bibfield  {title} {\bibinfo {title} {Goals and opportunities in quantum
  simulation},\ }\href {https://doi.org/10.1038/nphys2275} {\bibfield
  {journal} {\bibinfo  {journal} {Nat. Phys.}\ }\textbf {\bibinfo {volume}
  {8}},\ \bibinfo {pages} {264} (\bibinfo {year} {2012})}\BibitemShut {NoStop}%
\bibitem [{\citenamefont {Blatt}\ and\ \citenamefont
  {Roos}(2012)}]{blatt2012quantum}%
  \BibitemOpen
  \bibfield  {author} {\bibinfo {author} {\bibfnamefont {R.}~\bibnamefont
  {Blatt}}\ and\ \bibinfo {author} {\bibfnamefont {C.~F.}\ \bibnamefont
  {Roos}},\ }\bibfield  {title} {\bibinfo {title} {Quantum simulations with
  trapped ions},\ }\href {https://doi.org/10.1038/nphys2252} {\bibfield
  {journal} {\bibinfo  {journal} {Nat. Phys.}\ }\textbf {\bibinfo {volume}
  {8}},\ \bibinfo {pages} {277} (\bibinfo {year} {2012})}\BibitemShut {NoStop}%
\bibitem [{\citenamefont {Babbush}\ \emph {et~al.}(2023)\citenamefont
  {Babbush}, \citenamefont {Huggins}, \citenamefont {Berry}, \citenamefont
  {Ung}, \citenamefont {Zhao}, \citenamefont {Reichman}, \citenamefont {Neven},
  \citenamefont {Baczewski},\ and\ \citenamefont {Lee}}]{babbush2023quantum}%
  \BibitemOpen
  \bibfield  {author} {\bibinfo {author} {\bibfnamefont {R.}~\bibnamefont
  {Babbush}}, \bibinfo {author} {\bibfnamefont {W.~J.}\ \bibnamefont
  {Huggins}}, \bibinfo {author} {\bibfnamefont {D.~W.}\ \bibnamefont {Berry}},
  \bibinfo {author} {\bibfnamefont {S.~F.}\ \bibnamefont {Ung}}, \bibinfo
  {author} {\bibfnamefont {A.}~\bibnamefont {Zhao}}, \bibinfo {author}
  {\bibfnamefont {D.~R.}\ \bibnamefont {Reichman}}, \bibinfo {author}
  {\bibfnamefont {H.}~\bibnamefont {Neven}}, \bibinfo {author} {\bibfnamefont
  {A.~D.}\ \bibnamefont {Baczewski}},\ and\ \bibinfo {author} {\bibfnamefont
  {J.}~\bibnamefont {Lee}},\ }\href@noop {} {\bibinfo {title} {Quantum
  simulation of exact electron dynamics can be more efficient than classical
  mean-field methods}} (\bibinfo {year} {2023}),\ \Eprint
  {https://arxiv.org/abs/2301.01203} {arXiv:2301.01203} \BibitemShut {NoStop}%
\bibitem [{\citenamefont {Jin}\ \emph {et~al.}(2023)\citenamefont {Jin},
  \citenamefont {Liu}, \citenamefont {Li},\ and\ \citenamefont
  {Yu}}]{jin2023quantum}%
  \BibitemOpen
  \bibfield  {author} {\bibinfo {author} {\bibfnamefont {S.}~\bibnamefont
  {Jin}}, \bibinfo {author} {\bibfnamefont {N.}~\bibnamefont {Liu}}, \bibinfo
  {author} {\bibfnamefont {X.}~\bibnamefont {Li}},\ and\ \bibinfo {author}
  {\bibfnamefont {Y.}~\bibnamefont {Yu}},\ }\href@noop {} {\bibinfo {title}
  {Quantum simulation for quantum dynamics with artificial boundary
  conditions}} (\bibinfo {year} {2023}),\ \Eprint
  {https://arxiv.org/abs/2304.00667} {arXiv:2304.00667} \BibitemShut {NoStop}%
\bibitem [{\citenamefont {Zhang}\ \emph {et~al.}(2023)\citenamefont {Zhang},
  \citenamefont {Kim}, \citenamefont {Mark}, \citenamefont {Choi},\ and\
  \citenamefont {Painter}}]{zhang2023}%
  \BibitemOpen
  \bibfield  {author} {\bibinfo {author} {\bibfnamefont {X.}~\bibnamefont
  {Zhang}}, \bibinfo {author} {\bibfnamefont {E.}~\bibnamefont {Kim}}, \bibinfo
  {author} {\bibfnamefont {D.~K.}\ \bibnamefont {Mark}}, \bibinfo {author}
  {\bibfnamefont {S.}~\bibnamefont {Choi}},\ and\ \bibinfo {author}
  {\bibfnamefont {O.}~\bibnamefont {Painter}},\ }\bibfield  {title} {\bibinfo
  {title} {A superconducting quantum simulator based on a photonic-bandgap
  metamaterial},\ }\href {https://doi.org/10.1126/science.ade7651} {\bibfield
  {journal} {\bibinfo  {journal} {Science}\ }\textbf {\bibinfo {volume}
  {379}},\ \bibinfo {pages} {278} (\bibinfo {year} {2023})}\BibitemShut
  {NoStop}%
\bibitem [{\citenamefont {Sch\"{a}fer}\ \emph {et~al.}(2020)\citenamefont
  {Sch\"{a}fer}, \citenamefont {Fukuhara}, \citenamefont {Sugawa},
  \citenamefont {Takasu},\ and\ \citenamefont {Takahashi}}]{Schfer2020}%
  \BibitemOpen
  \bibfield  {author} {\bibinfo {author} {\bibfnamefont {F.}~\bibnamefont
  {Sch\"{a}fer}}, \bibinfo {author} {\bibfnamefont {T.}~\bibnamefont
  {Fukuhara}}, \bibinfo {author} {\bibfnamefont {S.}~\bibnamefont {Sugawa}},
  \bibinfo {author} {\bibfnamefont {Y.}~\bibnamefont {Takasu}},\ and\ \bibinfo
  {author} {\bibfnamefont {Y.}~\bibnamefont {Takahashi}},\ }\bibfield  {title}
  {\bibinfo {title} {Tools for quantum simulation with ultracold atoms in
  optical lattices},\ }\href {https://doi.org/10.1038/s42254-020-0195-3}
  {\bibfield  {journal} {\bibinfo  {journal} {Nat. Rev. Phys.}\ }\textbf
  {\bibinfo {volume} {2}},\ \bibinfo {pages} {411} (\bibinfo {year}
  {2020})}\BibitemShut {NoStop}%
\bibitem [{\citenamefont {Carrasco}\ \emph {et~al.}(2021)\citenamefont
  {Carrasco}, \citenamefont {Elben}, \citenamefont {Kokail}, \citenamefont
  {Kraus},\ and\ \citenamefont {Zoller}}]{quantum_verification}%
  \BibitemOpen
  \bibfield  {author} {\bibinfo {author} {\bibfnamefont {J.}~\bibnamefont
  {Carrasco}}, \bibinfo {author} {\bibfnamefont {A.}~\bibnamefont {Elben}},
  \bibinfo {author} {\bibfnamefont {C.}~\bibnamefont {Kokail}}, \bibinfo
  {author} {\bibfnamefont {B.}~\bibnamefont {Kraus}},\ and\ \bibinfo {author}
  {\bibfnamefont {P.}~\bibnamefont {Zoller}},\ }\bibfield  {title} {\bibinfo
  {title} {Theoretical and experimental perspectives of quantum verification},\
  }\href {https://doi.org/10.1103/PRXQuantum.2.010102} {\bibfield  {journal}
  {\bibinfo  {journal} {PRX Quantum}\ }\textbf {\bibinfo {volume} {2}},\
  \bibinfo {pages} {010102} (\bibinfo {year} {2021})}\BibitemShut {NoStop}%
\bibitem [{\citenamefont {Hauke}\ \emph {et~al.}(2012)\citenamefont {Hauke},
  \citenamefont {Cucchietti}, \citenamefont {Tagliacozzo}, \citenamefont
  {Deutsch},\ and\ \citenamefont {Lewenstein}}]{hauke2012}%
  \BibitemOpen
  \bibfield  {author} {\bibinfo {author} {\bibfnamefont {P.}~\bibnamefont
  {Hauke}}, \bibinfo {author} {\bibfnamefont {F.~M.}\ \bibnamefont
  {Cucchietti}}, \bibinfo {author} {\bibfnamefont {L.}~\bibnamefont
  {Tagliacozzo}}, \bibinfo {author} {\bibfnamefont {I.}~\bibnamefont
  {Deutsch}},\ and\ \bibinfo {author} {\bibfnamefont {M.}~\bibnamefont
  {Lewenstein}},\ }\bibfield  {title} {\bibinfo {title} {Can one trust quantum
  simulators?},\ }\href {https://doi.org/10.1088/0034-4885/75/8/082401}
  {\bibfield  {journal} {\bibinfo  {journal} {Rep. Prog. Phys.}\ }\textbf
  {\bibinfo {volume} {75}},\ \bibinfo {pages} {082401} (\bibinfo {year}
  {2012})}\BibitemShut {NoStop}%
\bibitem [{\citenamefont {Altman}\ \emph {et~al.}(2021)\citenamefont {Altman},
  \citenamefont {Brown}, \citenamefont {Carleo}, \citenamefont {Carr},
  \citenamefont {Demler}, \citenamefont {Chin}, \citenamefont {DeMarco},
  \citenamefont {Economou}, \citenamefont {Eriksson}, \citenamefont {Fu},
  \citenamefont {Greiner}, \citenamefont {Hazzard}, \citenamefont {Hulet},
  \citenamefont {Koll\'ar}, \citenamefont {Lev}, \citenamefont {Lukin},
  \citenamefont {Ma}, \citenamefont {Mi}, \citenamefont {Misra}, \citenamefont
  {Monroe}, \citenamefont {Murch}, \citenamefont {Nazario}, \citenamefont {Ni},
  \citenamefont {Potter}, \citenamefont {Roushan}, \citenamefont {Saffman},
  \citenamefont {Schleier-Smith}, \citenamefont {Siddiqi}, \citenamefont
  {Simmonds}, \citenamefont {Singh}, \citenamefont {Spielman}, \citenamefont
  {Temme}, \citenamefont {Weiss}, \citenamefont {Vu\ifmmode \check{c}\else
  \v{c}\fi{}kovi\ifmmode~\acute{c}\else \'{c}\fi{}}, \citenamefont
  {Vuleti\ifmmode~\acute{c}\else \'{c}\fi{}}, \citenamefont {Ye},\ and\
  \citenamefont {Zwierlein}}]{PRXQuantum.2.017003}%
  \BibitemOpen
  \bibfield  {author} {\bibinfo {author} {\bibfnamefont {E.}~\bibnamefont
  {Altman}}, \bibinfo {author} {\bibfnamefont {K.~R.}\ \bibnamefont {Brown}},
  \bibinfo {author} {\bibfnamefont {G.}~\bibnamefont {Carleo}}, \bibinfo
  {author} {\bibfnamefont {L.~D.}\ \bibnamefont {Carr}}, \bibinfo {author}
  {\bibfnamefont {E.}~\bibnamefont {Demler}}, \bibinfo {author} {\bibfnamefont
  {C.}~\bibnamefont {Chin}}, \bibinfo {author} {\bibfnamefont {B.}~\bibnamefont
  {DeMarco}}, \bibinfo {author} {\bibfnamefont {S.~E.}\ \bibnamefont
  {Economou}}, \bibinfo {author} {\bibfnamefont {M.~A.}\ \bibnamefont
  {Eriksson}}, \bibinfo {author} {\bibfnamefont {K.-M.~C.}\ \bibnamefont {Fu}},
  \bibinfo {author} {\bibfnamefont {M.}~\bibnamefont {Greiner}}, \bibinfo
  {author} {\bibfnamefont {K.~R.}\ \bibnamefont {Hazzard}}, \bibinfo {author}
  {\bibfnamefont {R.~G.}\ \bibnamefont {Hulet}}, \bibinfo {author}
  {\bibfnamefont {A.~J.}\ \bibnamefont {Koll\'ar}}, \bibinfo {author}
  {\bibfnamefont {B.~L.}\ \bibnamefont {Lev}}, \bibinfo {author} {\bibfnamefont
  {M.~D.}\ \bibnamefont {Lukin}}, \bibinfo {author} {\bibfnamefont
  {R.}~\bibnamefont {Ma}}, \bibinfo {author} {\bibfnamefont {X.}~\bibnamefont
  {Mi}}, \bibinfo {author} {\bibfnamefont {S.}~\bibnamefont {Misra}}, \bibinfo
  {author} {\bibfnamefont {C.}~\bibnamefont {Monroe}}, \bibinfo {author}
  {\bibfnamefont {K.}~\bibnamefont {Murch}}, \bibinfo {author} {\bibfnamefont
  {Z.}~\bibnamefont {Nazario}}, \bibinfo {author} {\bibfnamefont {K.-K.}\
  \bibnamefont {Ni}}, \bibinfo {author} {\bibfnamefont {A.~C.}\ \bibnamefont
  {Potter}}, \bibinfo {author} {\bibfnamefont {P.}~\bibnamefont {Roushan}},
  \bibinfo {author} {\bibfnamefont {M.}~\bibnamefont {Saffman}}, \bibinfo
  {author} {\bibfnamefont {M.}~\bibnamefont {Schleier-Smith}}, \bibinfo
  {author} {\bibfnamefont {I.}~\bibnamefont {Siddiqi}}, \bibinfo {author}
  {\bibfnamefont {R.}~\bibnamefont {Simmonds}}, \bibinfo {author}
  {\bibfnamefont {M.}~\bibnamefont {Singh}}, \bibinfo {author} {\bibfnamefont
  {I.}~\bibnamefont {Spielman}}, \bibinfo {author} {\bibfnamefont
  {K.}~\bibnamefont {Temme}}, \bibinfo {author} {\bibfnamefont {D.~S.}\
  \bibnamefont {Weiss}}, \bibinfo {author} {\bibfnamefont {J.}~\bibnamefont
  {Vu\ifmmode \check{c}\else \v{c}\fi{}kovi\ifmmode~\acute{c}\else
  \'{c}\fi{}}}, \bibinfo {author} {\bibfnamefont {V.}~\bibnamefont
  {Vuleti\ifmmode~\acute{c}\else \'{c}\fi{}}}, \bibinfo {author} {\bibfnamefont
  {J.}~\bibnamefont {Ye}},\ and\ \bibinfo {author} {\bibfnamefont
  {M.}~\bibnamefont {Zwierlein}},\ }\bibfield  {title} {\bibinfo {title}
  {Quantum simulators: Architectures and opportunities},\ }\href
  {https://doi.org/10.1103/PRXQuantum.2.017003} {\bibfield  {journal} {\bibinfo
   {journal} {PRX Quantum}\ }\textbf {\bibinfo {volume} {2}},\ \bibinfo {pages}
  {017003} (\bibinfo {year} {2021})}\BibitemShut {NoStop}%
\bibitem [{\citenamefont {Morgado}\ and\ \citenamefont
  {Whitlock}(2021)}]{Morgado2021}%
  \BibitemOpen
  \bibfield  {author} {\bibinfo {author} {\bibfnamefont {M.}~\bibnamefont
  {Morgado}}\ and\ \bibinfo {author} {\bibfnamefont {S.}~\bibnamefont
  {Whitlock}},\ }\bibfield  {title} {\bibinfo {title} {{Quantum simulation and
  computing with {Rydberg}-interacting qubits}},\ }\bibfield  {journal}
  {\bibinfo  {journal} {AVS Quantum Sci.}\ }\textbf {\bibinfo {volume} {3}},\
  \href {https://doi.org/10.1116/5.0036562} {10.1116/5.0036562} (\bibinfo
  {year} {2021}),\ \bibinfo {note} {023501}\BibitemShut {NoStop}%
\bibitem [{\citenamefont {Gebhart}\ \emph {et~al.}(2023)\citenamefont
  {Gebhart}, \citenamefont {Santagati}, \citenamefont {Gentile}, \citenamefont
  {Gauger}, \citenamefont {Craig}, \citenamefont {Ares}, \citenamefont
  {Banchi}, \citenamefont {Marquardt}, \citenamefont {Pezz{\`e}},\ and\
  \citenamefont {Bonato}}]{gebhart2023}%
  \BibitemOpen
  \bibfield  {author} {\bibinfo {author} {\bibfnamefont {V.}~\bibnamefont
  {Gebhart}}, \bibinfo {author} {\bibfnamefont {R.}~\bibnamefont {Santagati}},
  \bibinfo {author} {\bibfnamefont {A.~A.}\ \bibnamefont {Gentile}}, \bibinfo
  {author} {\bibfnamefont {E.~M.}\ \bibnamefont {Gauger}}, \bibinfo {author}
  {\bibfnamefont {D.}~\bibnamefont {Craig}}, \bibinfo {author} {\bibfnamefont
  {N.}~\bibnamefont {Ares}}, \bibinfo {author} {\bibfnamefont {L.}~\bibnamefont
  {Banchi}}, \bibinfo {author} {\bibfnamefont {F.}~\bibnamefont {Marquardt}},
  \bibinfo {author} {\bibfnamefont {L.}~\bibnamefont {Pezz{\`e}}},\ and\
  \bibinfo {author} {\bibfnamefont {C.}~\bibnamefont {Bonato}},\ }\bibfield
  {title} {\bibinfo {title} {Learning quantum systems},\ }\href
  {https://doi.org/10.1038/s42254-022-00552-1} {\bibfield  {journal} {\bibinfo
  {journal} {Nat. Rev. Phys.}\ }\textbf {\bibinfo {volume} {5}},\ \bibinfo
  {pages} {141} (\bibinfo {year} {2023})}\BibitemShut {NoStop}%
\bibitem [{\citenamefont {Braccia}\ \emph {et~al.}(2022)\citenamefont
  {Braccia}, \citenamefont {Banchi},\ and\ \citenamefont
  {Caruso}}]{braccia2022}%
  \BibitemOpen
  \bibfield  {author} {\bibinfo {author} {\bibfnamefont {P.}~\bibnamefont
  {Braccia}}, \bibinfo {author} {\bibfnamefont {L.}~\bibnamefont {Banchi}},\
  and\ \bibinfo {author} {\bibfnamefont {F.}~\bibnamefont {Caruso}},\
  }\bibfield  {title} {\bibinfo {title} {Quantum noise sensing by generating
  fake noise},\ }\href {https://doi.org/10.1103/PhysRevApplied.17.024002}
  {\bibfield  {journal} {\bibinfo  {journal} {Phys. Rev. Appl.}\ }\textbf
  {\bibinfo {volume} {17}},\ \bibinfo {pages} {024002} (\bibinfo {year}
  {2022})}\BibitemShut {NoStop}%
\bibitem [{\citenamefont {Han}\ \emph {et~al.}(2021)\citenamefont {Han},
  \citenamefont {Glaz}, \citenamefont {Haile},\ and\ \citenamefont
  {Lai}}]{han2021tomography}%
  \BibitemOpen
  \bibfield  {author} {\bibinfo {author} {\bibfnamefont {C.-D.}\ \bibnamefont
  {Han}}, \bibinfo {author} {\bibfnamefont {B.}~\bibnamefont {Glaz}}, \bibinfo
  {author} {\bibfnamefont {M.}~\bibnamefont {Haile}},\ and\ \bibinfo {author}
  {\bibfnamefont {Y.-C.}\ \bibnamefont {Lai}},\ }\bibfield  {title} {\bibinfo
  {title} {Tomography of time-dependent quantum {Hamiltonians} with machine
  learning},\ }\href {https://doi.org/10.1103/PhysRevA.104.062404} {\bibfield
  {journal} {\bibinfo  {journal} {Phys. Rev. A}\ }\textbf {\bibinfo {volume}
  {104}},\ \bibinfo {pages} {062404} (\bibinfo {year} {2021})}\BibitemShut
  {NoStop}%
\bibitem [{\citenamefont {Mohseni}\ \emph {et~al.}(2022)\citenamefont
  {Mohseni}, \citenamefont {F{\"o}sel}, \citenamefont {Guo}, \citenamefont
  {Navarrete-Benlloch},\ and\ \citenamefont {Marquardt}}]{mohseni2022deep}%
  \BibitemOpen
  \bibfield  {author} {\bibinfo {author} {\bibfnamefont {N.}~\bibnamefont
  {Mohseni}}, \bibinfo {author} {\bibfnamefont {T.}~\bibnamefont {F{\"o}sel}},
  \bibinfo {author} {\bibfnamefont {L.}~\bibnamefont {Guo}}, \bibinfo {author}
  {\bibfnamefont {C.}~\bibnamefont {Navarrete-Benlloch}},\ and\ \bibinfo
  {author} {\bibfnamefont {F.}~\bibnamefont {Marquardt}},\ }\bibfield  {title}
  {\bibinfo {title} {Deep learning of quantum many-body dynamics via random
  driving},\ }\href {https://doi.org/10.22331/q-2022-05-17-714} {\bibfield
  {journal} {\bibinfo  {journal} {Quantum}\ }\textbf {\bibinfo {volume} {6}},\
  \bibinfo {pages} {714} (\bibinfo {year} {2022})}\BibitemShut {NoStop}%
\bibitem [{\citenamefont {Genois}\ \emph {et~al.}(2021)\citenamefont {Genois},
  \citenamefont {Gross}, \citenamefont {Di~Paolo}, \citenamefont {Stevenson},
  \citenamefont {Koolstra}, \citenamefont {Hashim}, \citenamefont {Siddiqi},\
  and\ \citenamefont {Blais}}]{genois2021quantum}%
  \BibitemOpen
  \bibfield  {author} {\bibinfo {author} {\bibfnamefont {{\'E}.}~\bibnamefont
  {Genois}}, \bibinfo {author} {\bibfnamefont {J.~A.}\ \bibnamefont {Gross}},
  \bibinfo {author} {\bibfnamefont {A.}~\bibnamefont {Di~Paolo}}, \bibinfo
  {author} {\bibfnamefont {N.~J.}\ \bibnamefont {Stevenson}}, \bibinfo {author}
  {\bibfnamefont {G.}~\bibnamefont {Koolstra}}, \bibinfo {author}
  {\bibfnamefont {A.}~\bibnamefont {Hashim}}, \bibinfo {author} {\bibfnamefont
  {I.}~\bibnamefont {Siddiqi}},\ and\ \bibinfo {author} {\bibfnamefont
  {A.}~\bibnamefont {Blais}},\ }\bibfield  {title} {\bibinfo {title}
  {Quantum-tailored machine-learning characterization of a superconducting
  qubit},\ }\href {https://doi.org/10.1103/PRXQuantum.2.040355} {\bibfield
  {journal} {\bibinfo  {journal} {PRX Quantum}\ }\textbf {\bibinfo {volume}
  {2}},\ \bibinfo {pages} {040355} (\bibinfo {year} {2021})}\BibitemShut
  {NoStop}%
\bibitem [{\citenamefont {Wilde}\ \emph {et~al.}(2022)\citenamefont {Wilde},
  \citenamefont {Kshetrimayum}, \citenamefont {Roth}, \citenamefont
  {Hangleiter}, \citenamefont {Sweke},\ and\ \citenamefont
  {Eisert}}]{Wilde2022}%
  \BibitemOpen
  \bibfield  {author} {\bibinfo {author} {\bibfnamefont {F.}~\bibnamefont
  {Wilde}}, \bibinfo {author} {\bibfnamefont {A.}~\bibnamefont {Kshetrimayum}},
  \bibinfo {author} {\bibfnamefont {I.}~\bibnamefont {Roth}}, \bibinfo {author}
  {\bibfnamefont {D.}~\bibnamefont {Hangleiter}}, \bibinfo {author}
  {\bibfnamefont {R.}~\bibnamefont {Sweke}},\ and\ \bibinfo {author}
  {\bibfnamefont {J.}~\bibnamefont {Eisert}},\ }\href@noop {} {\bibinfo {title}
  {Scalably learning quantum many-body {Hamiltonians} from dynamical data}}
  (\bibinfo {year} {2022}),\ \Eprint {https://arxiv.org/abs/2209.14328}
  {arXiv:2209.14328} \BibitemShut {NoStop}%
\bibitem [{\citenamefont {Bairey}\ \emph {et~al.}(2019)\citenamefont {Bairey},
  \citenamefont {Arad},\ and\ \citenamefont {Lindner}}]{Bairey2019}%
  \BibitemOpen
  \bibfield  {author} {\bibinfo {author} {\bibfnamefont {E.}~\bibnamefont
  {Bairey}}, \bibinfo {author} {\bibfnamefont {I.}~\bibnamefont {Arad}},\ and\
  \bibinfo {author} {\bibfnamefont {N.~H.}\ \bibnamefont {Lindner}},\
  }\bibfield  {title} {\bibinfo {title} {Learning a local {Hamiltonian} from
  local measurements},\ }\href {https://doi.org/10.1103/PhysRevLett.122.020504}
  {\bibfield  {journal} {\bibinfo  {journal} {Phys. Rev. Lett.}\ }\textbf
  {\bibinfo {volume} {122}},\ \bibinfo {pages} {020504} (\bibinfo {year}
  {2019})}\BibitemShut {NoStop}%
\bibitem [{\citenamefont {Di~Franco}\ \emph {et~al.}(2009)\citenamefont
  {Di~Franco}, \citenamefont {Paternostro},\ and\ \citenamefont
  {Kim}}]{di2009hamiltonian}%
  \BibitemOpen
  \bibfield  {author} {\bibinfo {author} {\bibfnamefont {C.}~\bibnamefont
  {Di~Franco}}, \bibinfo {author} {\bibfnamefont {M.}~\bibnamefont
  {Paternostro}},\ and\ \bibinfo {author} {\bibfnamefont {M.}~\bibnamefont
  {Kim}},\ }\bibfield  {title} {\bibinfo {title} {Hamiltonian tomography in an
  access-limited setting without state initialization},\ }\href
  {https://doi.org/10.1103/PhysRevLett.102.187203} {\bibfield  {journal}
  {\bibinfo  {journal} {Phys. Rev. Lett.}\ }\textbf {\bibinfo {volume} {102}},\
  \bibinfo {pages} {187203} (\bibinfo {year} {2009})}\BibitemShut {NoStop}%
\bibitem [{\citenamefont {Cole}\ \emph {et~al.}(2005)\citenamefont {Cole},
  \citenamefont {Schirmer}, \citenamefont {Greentree}, \citenamefont {Wellard},
  \citenamefont {Oi},\ and\ \citenamefont {Hollenberg}}]{cole2005identifying}%
  \BibitemOpen
  \bibfield  {author} {\bibinfo {author} {\bibfnamefont {J.~H.}\ \bibnamefont
  {Cole}}, \bibinfo {author} {\bibfnamefont {S.~G.}\ \bibnamefont {Schirmer}},
  \bibinfo {author} {\bibfnamefont {A.~D.}\ \bibnamefont {Greentree}}, \bibinfo
  {author} {\bibfnamefont {C.~J.}\ \bibnamefont {Wellard}}, \bibinfo {author}
  {\bibfnamefont {D.~K.}\ \bibnamefont {Oi}},\ and\ \bibinfo {author}
  {\bibfnamefont {L.~C.}\ \bibnamefont {Hollenberg}},\ }\bibfield  {title}
  {\bibinfo {title} {Identifying an experimental two-state {Hamiltonian} to
  arbitrary accuracy},\ }\href {https://doi.org/10.1103/PhysRevA.71.062312}
  {\bibfield  {journal} {\bibinfo  {journal} {Phys. Rev. A}\ }\textbf {\bibinfo
  {volume} {71}},\ \bibinfo {pages} {062312} (\bibinfo {year}
  {2005})}\BibitemShut {NoStop}%
\bibitem [{\citenamefont {Devitt}\ \emph {et~al.}(2006)\citenamefont {Devitt},
  \citenamefont {Cole},\ and\ \citenamefont {Hollenberg}}]{devitt2006scheme}%
  \BibitemOpen
  \bibfield  {author} {\bibinfo {author} {\bibfnamefont {S.~J.}\ \bibnamefont
  {Devitt}}, \bibinfo {author} {\bibfnamefont {J.~H.}\ \bibnamefont {Cole}},\
  and\ \bibinfo {author} {\bibfnamefont {L.~C.}\ \bibnamefont {Hollenberg}},\
  }\bibfield  {title} {\bibinfo {title} {Scheme for direct measurement of a
  general two-qubit {Hamiltonian}},\ }\href
  {https://doi.org/10.1103/PhysRevA.73.052317} {\bibfield  {journal} {\bibinfo
  {journal} {Phys. Rev. A}\ }\textbf {\bibinfo {volume} {73}},\ \bibinfo
  {pages} {052317} (\bibinfo {year} {2006})}\BibitemShut {NoStop}%
\bibitem [{\citenamefont {Wang}\ \emph {et~al.}(2017)\citenamefont {Wang},
  \citenamefont {Paesani}, \citenamefont {Santagati}, \citenamefont {Knauer},
  \citenamefont {Gentile}, \citenamefont {Wiebe}, \citenamefont {Petruzzella},
  \citenamefont {O’Brien}, \citenamefont {Rarity}, \citenamefont {Laing}
  \emph {et~al.}}]{wang2017experimental}%
  \BibitemOpen
  \bibfield  {author} {\bibinfo {author} {\bibfnamefont {J.}~\bibnamefont
  {Wang}}, \bibinfo {author} {\bibfnamefont {S.}~\bibnamefont {Paesani}},
  \bibinfo {author} {\bibfnamefont {R.}~\bibnamefont {Santagati}}, \bibinfo
  {author} {\bibfnamefont {S.}~\bibnamefont {Knauer}}, \bibinfo {author}
  {\bibfnamefont {A.~A.}\ \bibnamefont {Gentile}}, \bibinfo {author}
  {\bibfnamefont {N.}~\bibnamefont {Wiebe}}, \bibinfo {author} {\bibfnamefont
  {M.}~\bibnamefont {Petruzzella}}, \bibinfo {author} {\bibfnamefont {J.~L.}\
  \bibnamefont {O’Brien}}, \bibinfo {author} {\bibfnamefont {J.~G.}\
  \bibnamefont {Rarity}}, \bibinfo {author} {\bibfnamefont {A.}~\bibnamefont
  {Laing}}, \emph {et~al.},\ }\bibfield  {title} {\bibinfo {title}
  {Experimental quantum {Hamiltonian} learning},\ }\href
  {https://doi.org/10.1038/nphys4074} {\bibfield  {journal} {\bibinfo
  {journal} {Nat. Phys.}\ }\textbf {\bibinfo {volume} {13}},\ \bibinfo {pages}
  {551} (\bibinfo {year} {2017})}\BibitemShut {NoStop}%
\bibitem [{\citenamefont {Gentile}\ \emph {et~al.}(2021)\citenamefont
  {Gentile}, \citenamefont {Flynn}, \citenamefont {Knauer}, \citenamefont
  {Wiebe}, \citenamefont {Paesani}, \citenamefont {Granade}, \citenamefont
  {Rarity}, \citenamefont {Santagati},\ and\ \citenamefont
  {Laing}}]{gentile2021learning}%
  \BibitemOpen
  \bibfield  {author} {\bibinfo {author} {\bibfnamefont {A.~A.}\ \bibnamefont
  {Gentile}}, \bibinfo {author} {\bibfnamefont {B.}~\bibnamefont {Flynn}},
  \bibinfo {author} {\bibfnamefont {S.}~\bibnamefont {Knauer}}, \bibinfo
  {author} {\bibfnamefont {N.}~\bibnamefont {Wiebe}}, \bibinfo {author}
  {\bibfnamefont {S.}~\bibnamefont {Paesani}}, \bibinfo {author} {\bibfnamefont
  {C.~E.}\ \bibnamefont {Granade}}, \bibinfo {author} {\bibfnamefont {J.~G.}\
  \bibnamefont {Rarity}}, \bibinfo {author} {\bibfnamefont {R.}~\bibnamefont
  {Santagati}},\ and\ \bibinfo {author} {\bibfnamefont {A.}~\bibnamefont
  {Laing}},\ }\bibfield  {title} {\bibinfo {title} {Learning models of quantum
  systems from experiments},\ }\href
  {https://doi.org/10.1038/s41567-021-01201-7} {\bibfield  {journal} {\bibinfo
  {journal} {Nat. Phys.}\ }\textbf {\bibinfo {volume} {17}},\ \bibinfo {pages}
  {837} (\bibinfo {year} {2021})}\BibitemShut {NoStop}%
\bibitem [{\citenamefont {Nakajima}(1958)}]{nakajima1958}%
  \BibitemOpen
  \bibfield  {author} {\bibinfo {author} {\bibfnamefont {S.}~\bibnamefont
  {Nakajima}},\ }\bibfield  {title} {\bibinfo {title} {On quantum theory of
  transport phenomena: steady diffusion},\ }\href
  {https://doi.org/10.1143/PTP.20.948} {\bibfield  {journal} {\bibinfo
  {journal} {Prog. Theor. Phys.}\ }\textbf {\bibinfo {volume} {20}},\ \bibinfo
  {pages} {948} (\bibinfo {year} {1958})}\BibitemShut {NoStop}%
\bibitem [{\citenamefont {Zwanzig}(1960)}]{zwanzig1960}%
  \BibitemOpen
  \bibfield  {author} {\bibinfo {author} {\bibfnamefont {R.}~\bibnamefont
  {Zwanzig}},\ }\bibfield  {title} {\bibinfo {title} {Ensemble method in the
  theory of irreversibility},\ }\href {https://doi.org/10.1063/1.1731409}
  {\bibfield  {journal} {\bibinfo  {journal} {J. Chem. Phys.}\ }\textbf
  {\bibinfo {volume} {33}},\ \bibinfo {pages} {1338} (\bibinfo {year}
  {1960})}\BibitemShut {NoStop}%
\bibitem [{\citenamefont {Cerrillo}\ and\ \citenamefont
  {Cao}(2014)}]{cerrillo2014}%
  \BibitemOpen
  \bibfield  {author} {\bibinfo {author} {\bibfnamefont {J.}~\bibnamefont
  {Cerrillo}}\ and\ \bibinfo {author} {\bibfnamefont {J.}~\bibnamefont {Cao}},\
  }\bibfield  {title} {\bibinfo {title} {{Non-Markovian} dynamical maps:
  numerical processing of open quantum trajectories},\ }\href
  {https://doi.org/10.1103/PhysRevLett.112.110401} {\bibfield  {journal}
  {\bibinfo  {journal} {Phys. Rev. Lett.}\ }\textbf {\bibinfo {volume} {112}},\
  \bibinfo {pages} {110401} (\bibinfo {year} {2014})}\BibitemShut {NoStop}%
\bibitem [{\citenamefont {Gelzinis}\ \emph {et~al.}(2017)\citenamefont
  {Gelzinis}, \citenamefont {Rybakovas},\ and\ \citenamefont
  {Valkunas}}]{gelzinis2017}%
  \BibitemOpen
  \bibfield  {author} {\bibinfo {author} {\bibfnamefont {A.}~\bibnamefont
  {Gelzinis}}, \bibinfo {author} {\bibfnamefont {E.}~\bibnamefont
  {Rybakovas}},\ and\ \bibinfo {author} {\bibfnamefont {L.}~\bibnamefont
  {Valkunas}},\ }\bibfield  {title} {\bibinfo {title} {Applicability of
  transfer tensor method for open quantum system dynamics},\ }\href
  {https://doi.org/10.1063/1.5009086} {\bibfield  {journal} {\bibinfo
  {journal} {J. Chem. Phys.}\ }\textbf {\bibinfo {volume} {147}},\ \bibinfo
  {pages} {234108} (\bibinfo {year} {2017})}\BibitemShut {NoStop}%
\bibitem [{\citenamefont {Pollock}\ and\ \citenamefont
  {Modi}(2018)}]{pollock2018}%
  \BibitemOpen
  \bibfield  {author} {\bibinfo {author} {\bibfnamefont {F.~A.}\ \bibnamefont
  {Pollock}}\ and\ \bibinfo {author} {\bibfnamefont {K.}~\bibnamefont {Modi}},\
  }\bibfield  {title} {\bibinfo {title} {Tomographically reconstructed master
  equations for any open quantum dynamics},\ }\href
  {https://doi.org/10.22331/q-2018-07-11-76} {\bibfield  {journal} {\bibinfo
  {journal} {Quantum}\ }\textbf {\bibinfo {volume} {2}},\ \bibinfo {pages} {76}
  (\bibinfo {year} {2018})}\BibitemShut {NoStop}%
\bibitem [{\citenamefont {Banchi}\ \emph {et~al.}(2018)\citenamefont {Banchi},
  \citenamefont {Grant}, \citenamefont {Rocchetto},\ and\ \citenamefont
  {Severini}}]{banchi2018modelling}%
  \BibitemOpen
  \bibfield  {author} {\bibinfo {author} {\bibfnamefont {L.}~\bibnamefont
  {Banchi}}, \bibinfo {author} {\bibfnamefont {E.}~\bibnamefont {Grant}},
  \bibinfo {author} {\bibfnamefont {A.}~\bibnamefont {Rocchetto}},\ and\
  \bibinfo {author} {\bibfnamefont {S.}~\bibnamefont {Severini}},\ }\bibfield
  {title} {\bibinfo {title} {Modelling non-{Markovian} quantum processes with
  recurrent neural networks},\ }\href
  {https://doi.org/10.1088/1367-2630/aaf749} {\bibfield  {journal} {\bibinfo
  {journal} {New J. Phys.}\ }\textbf {\bibinfo {volume} {20}},\ \bibinfo
  {pages} {123030} (\bibinfo {year} {2018})}\BibitemShut {NoStop}%
\bibitem [{\citenamefont {Link}\ \emph {et~al.}(2023)\citenamefont {Link},
  \citenamefont {Gao}, \citenamefont {Kell}, \citenamefont {Breyer},
  \citenamefont {Eberz}, \citenamefont {Rauf},\ and\ \citenamefont
  {K\"ohl}}]{link2023}%
  \BibitemOpen
  \bibfield  {author} {\bibinfo {author} {\bibfnamefont {M.}~\bibnamefont
  {Link}}, \bibinfo {author} {\bibfnamefont {K.}~\bibnamefont {Gao}}, \bibinfo
  {author} {\bibfnamefont {A.}~\bibnamefont {Kell}}, \bibinfo {author}
  {\bibfnamefont {M.}~\bibnamefont {Breyer}}, \bibinfo {author} {\bibfnamefont
  {D.}~\bibnamefont {Eberz}}, \bibinfo {author} {\bibfnamefont
  {B.}~\bibnamefont {Rauf}},\ and\ \bibinfo {author} {\bibfnamefont
  {M.}~\bibnamefont {K\"ohl}},\ }\bibfield  {title} {\bibinfo {title} {Machine
  learning the phase diagram of a strongly interacting fermi gas},\ }\href
  {https://doi.org/10.1103/PhysRevLett.130.203401} {\bibfield  {journal}
  {\bibinfo  {journal} {Phys. Rev. Lett.}\ }\textbf {\bibinfo {volume} {130}},\
  \bibinfo {pages} {203401} (\bibinfo {year} {2023})}\BibitemShut {NoStop}%
\bibitem [{\citenamefont {Wu}\ \emph {et~al.}(2021)\citenamefont {Wu},
  \citenamefont {Liang}, \citenamefont {Tian}, \citenamefont {Yang},
  \citenamefont {Chen}, \citenamefont {Liu}, \citenamefont {Tey},\ and\
  \citenamefont {You}}]{Wu_2021}%
  \BibitemOpen
  \bibfield  {author} {\bibinfo {author} {\bibfnamefont {X.}~\bibnamefont
  {Wu}}, \bibinfo {author} {\bibfnamefont {X.}~\bibnamefont {Liang}}, \bibinfo
  {author} {\bibfnamefont {Y.}~\bibnamefont {Tian}}, \bibinfo {author}
  {\bibfnamefont {F.}~\bibnamefont {Yang}}, \bibinfo {author} {\bibfnamefont
  {C.}~\bibnamefont {Chen}}, \bibinfo {author} {\bibfnamefont {Y.-C.}\
  \bibnamefont {Liu}}, \bibinfo {author} {\bibfnamefont {M.~K.}\ \bibnamefont
  {Tey}},\ and\ \bibinfo {author} {\bibfnamefont {L.}~\bibnamefont {You}},\
  }\bibfield  {title} {\bibinfo {title} {A concise review of {Rydberg} atom
  based quantum computation and quantum simulation*},\ }\href
  {https://doi.org/10.1088/1674-1056/abd76f} {\bibfield  {journal} {\bibinfo
  {journal} {Chin. Phys. B}\ }\textbf {\bibinfo {volume} {30}},\ \bibinfo
  {pages} {020305} (\bibinfo {year} {2021})}\BibitemShut {NoStop}%
\bibitem [{\citenamefont {Saffman}\ \emph {et~al.}(2010)\citenamefont
  {Saffman}, \citenamefont {Walker},\ and\ \citenamefont
  {M\o{}lmer}}]{Saffman2010}%
  \BibitemOpen
  \bibfield  {author} {\bibinfo {author} {\bibfnamefont {M.}~\bibnamefont
  {Saffman}}, \bibinfo {author} {\bibfnamefont {T.~G.}\ \bibnamefont
  {Walker}},\ and\ \bibinfo {author} {\bibfnamefont {K.}~\bibnamefont
  {M\o{}lmer}},\ }\bibfield  {title} {\bibinfo {title} {Quantum information
  with {Rydberg} atoms},\ }\href {https://doi.org/10.1103/RevModPhys.82.2313}
  {\bibfield  {journal} {\bibinfo  {journal} {Rev. Mod. Phys.}\ }\textbf
  {\bibinfo {volume} {82}},\ \bibinfo {pages} {2313} (\bibinfo {year}
  {2010})}\BibitemShut {NoStop}%
\bibitem [{\citenamefont {Mazza}\ \emph {et~al.}(2021)\citenamefont {Mazza},
  \citenamefont {Zietlow}, \citenamefont {Carollo}, \citenamefont
  {Andergassen}, \citenamefont {Martius},\ and\ \citenamefont
  {Lesanovsky}}]{Mazza2021}%
  \BibitemOpen
  \bibfield  {author} {\bibinfo {author} {\bibfnamefont {P.~P.}\ \bibnamefont
  {Mazza}}, \bibinfo {author} {\bibfnamefont {D.}~\bibnamefont {Zietlow}},
  \bibinfo {author} {\bibfnamefont {F.}~\bibnamefont {Carollo}}, \bibinfo
  {author} {\bibfnamefont {S.}~\bibnamefont {Andergassen}}, \bibinfo {author}
  {\bibfnamefont {G.}~\bibnamefont {Martius}},\ and\ \bibinfo {author}
  {\bibfnamefont {I.}~\bibnamefont {Lesanovsky}},\ }\bibfield  {title}
  {\bibinfo {title} {Machine learning time-local generators of open quantum
  dynamics},\ }\href {https://doi.org/10.1103/PhysRevResearch.3.023084}
  {\bibfield  {journal} {\bibinfo  {journal} {Phys. Rev. Res.}\ }\textbf
  {\bibinfo {volume} {3}},\ \bibinfo {pages} {023084} (\bibinfo {year}
  {2021})}\BibitemShut {NoStop}%
\bibitem [{\citenamefont {Carnazza}\ \emph {et~al.}(2022)\citenamefont
  {Carnazza}, \citenamefont {Carollo}, \citenamefont {Zietlow}, \citenamefont
  {Andergassen}, \citenamefont {Martius},\ and\ \citenamefont
  {Lesanovsky}}]{Carnazza2022}%
  \BibitemOpen
  \bibfield  {author} {\bibinfo {author} {\bibfnamefont {F.}~\bibnamefont
  {Carnazza}}, \bibinfo {author} {\bibfnamefont {F.}~\bibnamefont {Carollo}},
  \bibinfo {author} {\bibfnamefont {D.}~\bibnamefont {Zietlow}}, \bibinfo
  {author} {\bibfnamefont {S.}~\bibnamefont {Andergassen}}, \bibinfo {author}
  {\bibfnamefont {G.}~\bibnamefont {Martius}},\ and\ \bibinfo {author}
  {\bibfnamefont {I.}~\bibnamefont {Lesanovsky}},\ }\bibfield  {title}
  {\bibinfo {title} {Inferring {Markovian} quantum master equations of few-body
  observables in interacting spin chains},\ }\href
  {https://doi.org/10.1088/1367-2630/ac7df6} {\bibfield  {journal} {\bibinfo
  {journal} {New J. Phys.}\ }\textbf {\bibinfo {volume} {24}},\ \bibinfo
  {pages} {073033} (\bibinfo {year} {2022})}\BibitemShut {NoStop}%
\bibitem [{\citenamefont {Breuer}\ and\ \citenamefont
  {Petruccione}(2002)}]{BRE02}%
  \BibitemOpen
  \bibfield  {author} {\bibinfo {author} {\bibfnamefont {H.~P.}\ \bibnamefont
  {Breuer}}\ and\ \bibinfo {author} {\bibfnamefont {F.}~\bibnamefont
  {Petruccione}},\ }\href@noop {} {\emph {\bibinfo {title} {The theory of open
  quantum systems}}}\ (\bibinfo  {publisher} {Oxford University Press},\
  \bibinfo {address} {Great Clarendon Street},\ \bibinfo {year}
  {2002})\BibitemShut {NoStop}%
\bibitem [{\citenamefont {Gorini}\ \emph {et~al.}(1976)\citenamefont {Gorini},
  \citenamefont {Kossakowski},\ and\ \citenamefont {Sudarshan}}]{Gorini1976}%
  \BibitemOpen
  \bibfield  {author} {\bibinfo {author} {\bibfnamefont {V.}~\bibnamefont
  {Gorini}}, \bibinfo {author} {\bibfnamefont {A.}~\bibnamefont
  {Kossakowski}},\ and\ \bibinfo {author} {\bibfnamefont {E.~C.~G.}\
  \bibnamefont {Sudarshan}},\ }\bibfield  {title} {\bibinfo {title} {Completely
  positive dynamical semigroups of {N-level} systems},\ }\href
  {https://doi.org/10.1063/1.522979} {\bibfield  {journal} {\bibinfo  {journal}
  {J. Math. Phys.}\ }\textbf {\bibinfo {volume} {17}},\ \bibinfo {pages} {821}
  (\bibinfo {year} {1976})}\BibitemShut {NoStop}%
\bibitem [{\citenamefont {Lindblad}(1976)}]{Lindblad1976}%
  \BibitemOpen
  \bibfield  {author} {\bibinfo {author} {\bibfnamefont {G.}~\bibnamefont
  {Lindblad}},\ }\bibfield  {title} {\bibinfo {title} {On the generators of
  quantum dynamical semigroups},\ }\href {https://doi.org/10.1007/bf01608499}
  {\bibfield  {journal} {\bibinfo  {journal} {Commun. Math. Phys.}\ }\textbf
  {\bibinfo {volume} {48}},\ \bibinfo {pages} {119} (\bibinfo {year}
  {1976})}\BibitemShut {NoStop}%
\bibitem [{\citenamefont {Vidal}(2004)}]{Vidal2004}%
  \BibitemOpen
  \bibfield  {author} {\bibinfo {author} {\bibfnamefont {G.}~\bibnamefont
  {Vidal}},\ }\bibfield  {title} {\bibinfo {title} {Efficient simulation of
  one-dimensional quantum many-body systems},\ }\href
  {https://doi.org/10.1103/PhysRevLett.93.040502} {\bibfield  {journal}
  {\bibinfo  {journal} {Phys. Rev. Lett.}\ }\textbf {\bibinfo {volume} {93}},\
  \bibinfo {pages} {040502} (\bibinfo {year} {2004})}\BibitemShut {NoStop}%
\bibitem [{\citenamefont {Zwolak}\ and\ \citenamefont
  {Vidal}(2004)}]{Zwolak2004}%
  \BibitemOpen
  \bibfield  {author} {\bibinfo {author} {\bibfnamefont {M.}~\bibnamefont
  {Zwolak}}\ and\ \bibinfo {author} {\bibfnamefont {G.}~\bibnamefont {Vidal}},\
  }\bibfield  {title} {\bibinfo {title} {Mixed-state dynamics in
  one-dimensional quantum lattice systems: a time-dependent superoperator
  renormalization algorithm},\ }\href
  {https://doi.org/10.1103/PhysRevLett.93.207205} {\bibfield  {journal}
  {\bibinfo  {journal} {Phys. Rev. Lett.}\ }\textbf {\bibinfo {volume} {93}},\
  \bibinfo {pages} {207205} (\bibinfo {year} {2004})}\BibitemShut {NoStop}%
\bibitem [{\citenamefont {Gray}(2018)}]{gray2018quimb}%
  \BibitemOpen
  \bibfield  {author} {\bibinfo {author} {\bibfnamefont {J.}~\bibnamefont
  {Gray}},\ }\bibfield  {title} {\bibinfo {title} {quimb: a python library for
  quantum information and many-body calculations},\ }\href@noop {} {\bibfield
  {journal} {\bibinfo  {journal} {Journal of Open Source Software}\ }\textbf
  {\bibinfo {volume} {3}},\ \bibinfo {pages} {819} (\bibinfo {year}
  {2018})}\BibitemShut {NoStop}%
\bibitem [{\citenamefont {Byrd}\ and\ \citenamefont
  {Khaneja}(2003)}]{Byrd2003}%
  \BibitemOpen
  \bibfield  {author} {\bibinfo {author} {\bibfnamefont {M.~S.}\ \bibnamefont
  {Byrd}}\ and\ \bibinfo {author} {\bibfnamefont {N.}~\bibnamefont {Khaneja}},\
  }\bibfield  {title} {\bibinfo {title} {Characterization of the positivity of
  the density matrix in terms of the coherence vector representation},\ }\href
  {https://doi.org/10.1103/PhysRevA.68.062322} {\bibfield  {journal} {\bibinfo
  {journal} {Phys. Rev. A}\ }\textbf {\bibinfo {volume} {68}},\ \bibinfo
  {pages} {062322} (\bibinfo {year} {2003})}\BibitemShut {NoStop}%
\bibitem [{SM()}]{SM}%
  \BibitemOpen
  \href@noop {} {}\bibinfo {note} {See Supplemental Material for
  details.}\BibitemShut {Stop}%
\bibitem [{\citenamefont {Imambi}\ \emph {et~al.}(2021)\citenamefont {Imambi},
  \citenamefont {Prakash},\ and\ \citenamefont
  {Kanagachidambaresan}}]{pytorch}%
  \BibitemOpen
  \bibfield  {author} {\bibinfo {author} {\bibfnamefont {S.}~\bibnamefont
  {Imambi}}, \bibinfo {author} {\bibfnamefont {K.~B.}\ \bibnamefont
  {Prakash}},\ and\ \bibinfo {author} {\bibfnamefont {G.}~\bibnamefont
  {Kanagachidambaresan}},\ }\bibfield  {title} {\bibinfo {title} {Pytorch},\
  }\href@noop {} {\bibfield  {journal} {\bibinfo  {journal} {Programming with
  {TensorFlow}: Solution for Edge Computing Applications}\ ,\ \bibinfo {pages}
  {87}} (\bibinfo {year} {2021})}\BibitemShut {NoStop}%
\bibitem [{\citenamefont {Kingma}\ and\ \citenamefont {Ba}(2014)}]{adam}%
  \BibitemOpen
  \bibfield  {author} {\bibinfo {author} {\bibfnamefont {D.~P.}\ \bibnamefont
  {Kingma}}\ and\ \bibinfo {author} {\bibfnamefont {J.}~\bibnamefont {Ba}},\
  }\href@noop {} {\bibinfo {title} {Adam: A method for stochastic
  optimization}} (\bibinfo {year} {2014}),\ \Eprint
  {https://arxiv.org/abs/1412.6980} {arXiv:1412.6980} \BibitemShut {NoStop}%
\bibitem [{Note1()}]{Note1}%
  \BibitemOpen
  \bibinfo {note} {The code for the generation of the artificial data and the
  training of the ML algorithm is made available at \protect \url
  {https://github.com/giovannicemin/lindblad_dynamics_approximator}.}\BibitemShut
  {Stop}%
\bibitem [{\citenamefont {Roos}\ \emph {et~al.}(2020)\citenamefont {Roos},
  \citenamefont {Cirac},\ and\ \citenamefont {Ba{\~n}uls}}]{Roos2020}%
  \BibitemOpen
  \bibfield  {author} {\bibinfo {author} {\bibfnamefont {J.}~\bibnamefont
  {Roos}}, \bibinfo {author} {\bibfnamefont {J.~I.}\ \bibnamefont {Cirac}},\
  and\ \bibinfo {author} {\bibfnamefont {M.~C.}\ \bibnamefont {Ba{\~n}uls}},\
  }\bibfield  {title} {\bibinfo {title} {{Markovianity} of an emitter coupled
  to a structured spin-chain bath},\ }\href
  {https://doi.org/10.1103/PhysRevA.101.042114} {\bibfield  {journal} {\bibinfo
   {journal} {Phys. Rev. A}\ }\textbf {\bibinfo {volume} {101}},\ \bibinfo
  {pages} {042114} (\bibinfo {year} {2020})}\BibitemShut {NoStop}%
\end{thebibliography}%
\newpage

\setcounter{equation}{0}
\setcounter{figure}{0}
\setcounter{table}{0}
\makeatletter
\renewcommand{\theequation}{S\arabic{equation}}
\renewcommand{\thefigure}{S\arabic{figure}}
\makeatletter

\onecolumngrid
\newpage

\setcounter{page}{1}
\begin{center}
{\Large SUPPLEMENTAL MATERIAL}
\end{center}
\begin{center}
\vspace{0.8cm}
{\Large Inferring interpretable dynamical generators of local quantum observables from projective measurements through machine learning}
\end{center}
\begin{center}
Giovanni Cemin,$^{1}$ Francesco Carnazza,$^{1}$ Sabine Andergassen,$^{2}$\\
Georg Martius,$^{3,4}$ Federico Carollo,$^{1}$ and Igor Lesanovsky$^{1,5}$
\end{center}
\begin{center}
$^1${\em Institut f\"ur Theoretische Physik, Universit\"at T\"ubingen,}\\
{\em Auf der Morgenstelle 14, 72076 T\"ubingen, Germany}\\
$^2${\em Institute for Solid State Physics and Institute of Information Systems Engineering, Vienna University of Technology, 1040 Vienna, Austria}\\
$^3${\em Max Planck Institute for Intelligent Systems, Max-Planck-Ring 4, 72076 Tübingen, Germany}\\

$^4$ {\em Wilhelm Schickard Institut für
Informatik, Maria-von-Linden-Straße 6
72076 T\"{u}bingen}\\
$^5${\em School of Physics and Astronomy and Centre for the Mathematics}\\
{\em and Theoretical Physics of Quantum Non-Equilibrium Systems,}\\
{\em  The University of Nottingham, Nottingham, NG7 2RD, United Kingdom}\\

\end{center}

\section*{I. Matrix representation of $\mathbf{L}$}
We here report a brief derivation of the explicit matrix elements appearing in eq. \eqref{eq_master_eq}. The stating point is the coherence vector $v_i = \Tr(F_i \rho_S)$. The time derivative is simply given by
\begin{equation}
    \frac{d}{dt} v_i (t) = \Tr \bigg( F_i \frac{d}{dt} \rho_S (t) \bigg) = \Tr( F_i \mathcal{L}[\rho_S(t)] ) = \Tr( \mathcal{L}^*[F_i] \rho_S (t) )
    \label{eq_S1}
\end{equation}
where in the second equality we used eq. \eqref{eq_master_eq_formal}. In the last equality $\mathcal{L}^*$ is the dual map of $\mathcal{L}$, i.e. the one evolving observables instead of states.
By expanding the dual map $\mathcal{L}^*[F_i] = \sum_{j=1}^{d^2} \Tr ( \mathcal{L}^*[F_i] F_j )F_j$, and substituting into eq. \eqref{eq_S1}, one obtains
\begin{equation}
     \frac{d}{dt} v_i (t) = \sum_{j=1}^{d^2} \Tr ( \mathcal{L}^*[F_i] F_j ) \Tr (F_j \rho_S ) = [\mathbf{Lv}(t)]_i \, ,
\end{equation}
where we defined the matrix $\mathbf{L}$ as $\mathbf{L}_{ij} \equiv \Tr( \mathcal{L}^*[F_i] F_j )$.
By exploiting the cyclic property of the trace, one can explicitly calculate $\mathcal{L}^*[F_i]$ starting from the known action of $\mathcal{L}[\rho]$. Explicitly we obtain
\begin{equation}
    \Tr (\mathcal{L}^*[F_k]F_l) = \Tr \bigg( i[H, F_k] F_l + \frac{1}{2} \sum_{i,j=2}^{d^2} c_{ij} ( F^{\dagger}_j[F_k,F_i]F_l + [F^{\dagger}_j,F_k]F_i F_l ) \bigg) \, .
\end{equation}
Comparing this result to eq. \eqref{eq_master_eq} we obtain
\begin{equation}
    \mathbf{H}_{kl} = i \Tr( [H,F_k]F_l) \, , \qquad \mathbf{D}_{kl} =  \frac{1}{2} \sum_{i,j=2}^{d^2} c_{ij} \Tr( F^{\dagger}_j[F_k,F_i]F_l + [F^{\dagger}_j,F_k]F_i F_l ) \, .
    \label{eq_S2}
\end{equation}

To simplify this expressions we use the structure constants $d_{ijk}$ and $f_{ijk}$, defined from the commutation and anticommutaton relations as
\begin{align}
    d_{ijk} = \frac{1}{4} \Tr ( \{ F_i, F_j\} F_k ) \, , \\
    f_{ijk} = -\frac{i}{4} \Tr( [F_i,F_j]F_k ) \, .
\end{align}
By substituting those definitions into eq. \eqref{eq_S2}, we obtain the following expressions for the matrix elements of $\mathbf{H}$ 
\begin{equation}
    \begin{aligned}
    \mathbf{H}_{ij} &= -4 \sum_{k=2}^{d^2} f_{ijk} \theta^H_k \, , \quad &i,j \in \{2,...,d^2\} \\
    \mathbf{H}_{i1} &= \mathbf{H}_{1i} = 0\, ,  \quad &i \in \{1,...,d^2\}
    \end{aligned}
    \label{eq_H_matrix}
\end{equation}
where $\theta^H_k$ provides the expansion of the Hamiltonian over the basis $F_i$, namely $H=\sum_{i=2}^{i=d^2} F_i \theta^H_i$. The matrix elements of $\mathbf{D}$ reads
\begin{equation}
    \begin{aligned}
    \mathbf{D}_{mn} &= -8 \sum_{i,j,k=2}^{d^2} f_{mik}( f_{njk} \Re (c)_{ij} + d_{njk} \Im (c)_{ij}) \, ,  &m,n \in \{2,...,d^2\} \\
    \mathbf{D}_{m1} &= 2 \sum_{i,j=2}^{d^2} f_{imj} \Im(c)_{ij} \, , \quad \mathbf{D}_{1m} = 0 \, ,
    &m \in \{1,...,d^2\}
    \end{aligned}
    \label{eq_D_matrix}
\end{equation}
where $c_{ij}$ is the so-called Kossakowski matrix, which we parametrized as $c=Z^\dagger Z$ with $Z=\theta^X+i\theta^Y$ and $\theta^X$,$\theta^Y$ real matrices.

Equations \eqref{eq_H_matrix} and \eqref{eq_D_matrix} are the matrix representation of the Lindblad operator $\mathcal{L}$ in terms of the vector $\omega_i$ and the Kossakowski matrix $c_{ij}$.

\section*{II. Details on the training procedure}
The loss used for the optimization of the LDA parameters is the mean-squared error function ($\mathbb{E}_{D}$ indicates average over the dataset)
\begin{equation}
    \mathrm{MSE}(\theta) = \mathbb{E}_{D} \Bigl[ || M(\theta, \tau)\mathbf{v}(0) - \mathbf{v}(\tau) ||^2 \Bigl] \, . 
\end{equation}
We additionally consider a regularization term on the parameters penalizing nonzero elements of $\theta$. The rationale behind this term is twofold: 
first, it yields a more stable training procedure, especially for small training datasets;
second, it keeps the learned generator as simple as possible for enhanced interpretability. The total loss function is thus 
\begin{equation}
    \mathrm{Loss}(\theta) = \mathrm{MSE}(\theta) + \alpha_{11} (|| \theta^X ||_1 + || \theta^Y||_1) + \alpha_{12} ||\theta^H||_1 \, ,
    \label{eq_loss}
\end{equation}
where $\|O\|_1=\sum_{i,j} |O_{ij}|$.
The LDA parameters $\theta$ are optimized by means of Adam optimization algorithm \cite{adam}, with a scheduled learning rate, where it decays by a factor of $0.05$ at 100th and 50th epochs from the final one.

\section*{III. Further results}
\begin{figure}[h]
    \centering
    \includegraphics[scale=0.5]{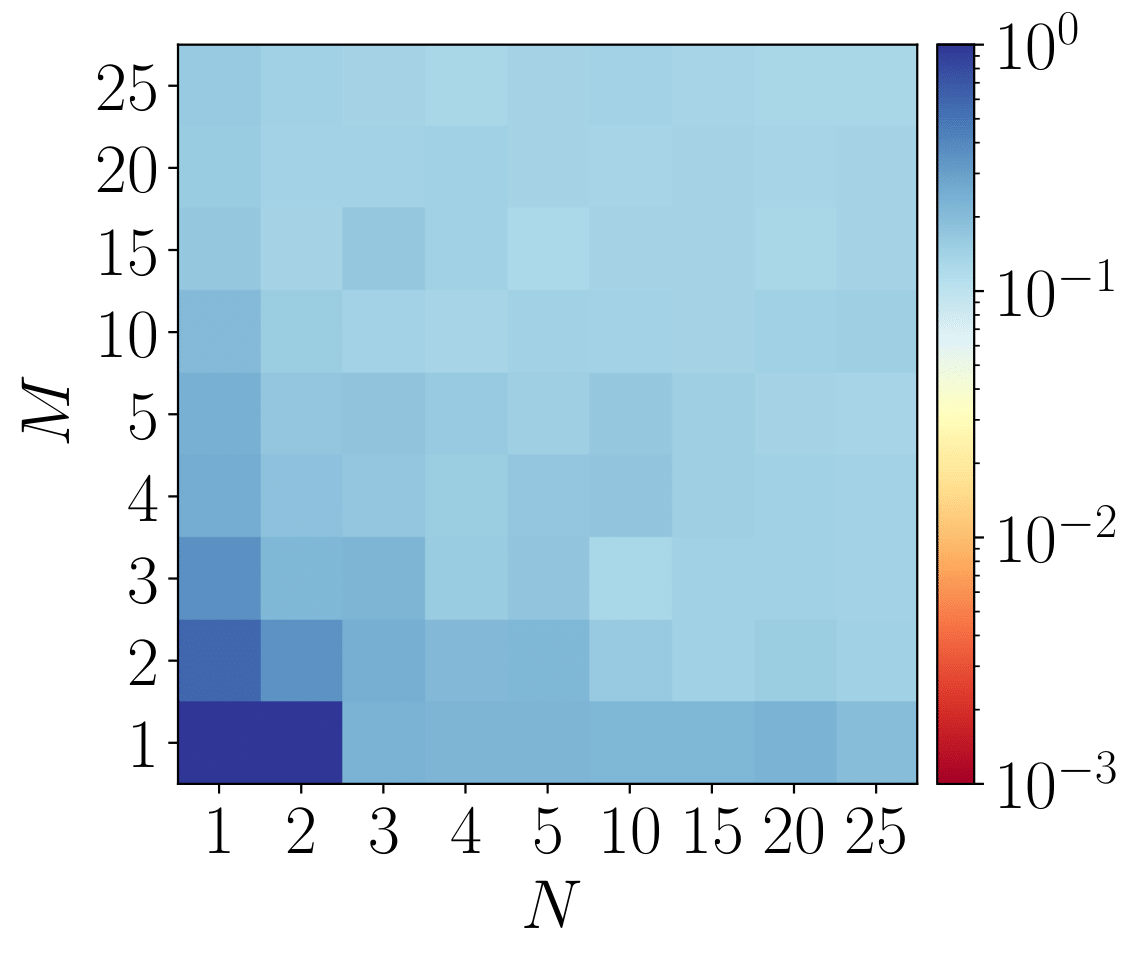}
    \caption{{\bf ML algorithm's performance on open system dynamics.} The figure reports the error $\epsilon(N,M)$ as in \eqref{eq_error_model} plotted for different values of $N$ and $M$, and for the $V=0.5\Omega$ spin chain.}
    \label{fig_case_05}
\end{figure}

In the main text, we investigate the behavior of the ML framework applied to spin chains having $V=0.1\Omega, 2\Omega$. Here we report the case of a spin chain having $V = 0.5 \Omega$. The results are shownin Figure \ref{fig_case_05}. The graph exhibits a similar behavior to the cases analyzed in the main text. However, there is a notable distinction: the model saturates at a higher error. This discrepancy arises from the stronger coupling between the system and the bath, rendering non-Markovian effects non-negligible.

\section*{IV. Additional plots and physical ``read out"}
For completeness, we report here the plots of all $15$ non-trivial components of the coherence vector $v_i$. The plots represent the exact time evolution (black solid line) and the dynamics predicted by the ML algorithm (red dashed line).
\begin{figure}[t!]
    \centering
    \includegraphics[scale=0.45]{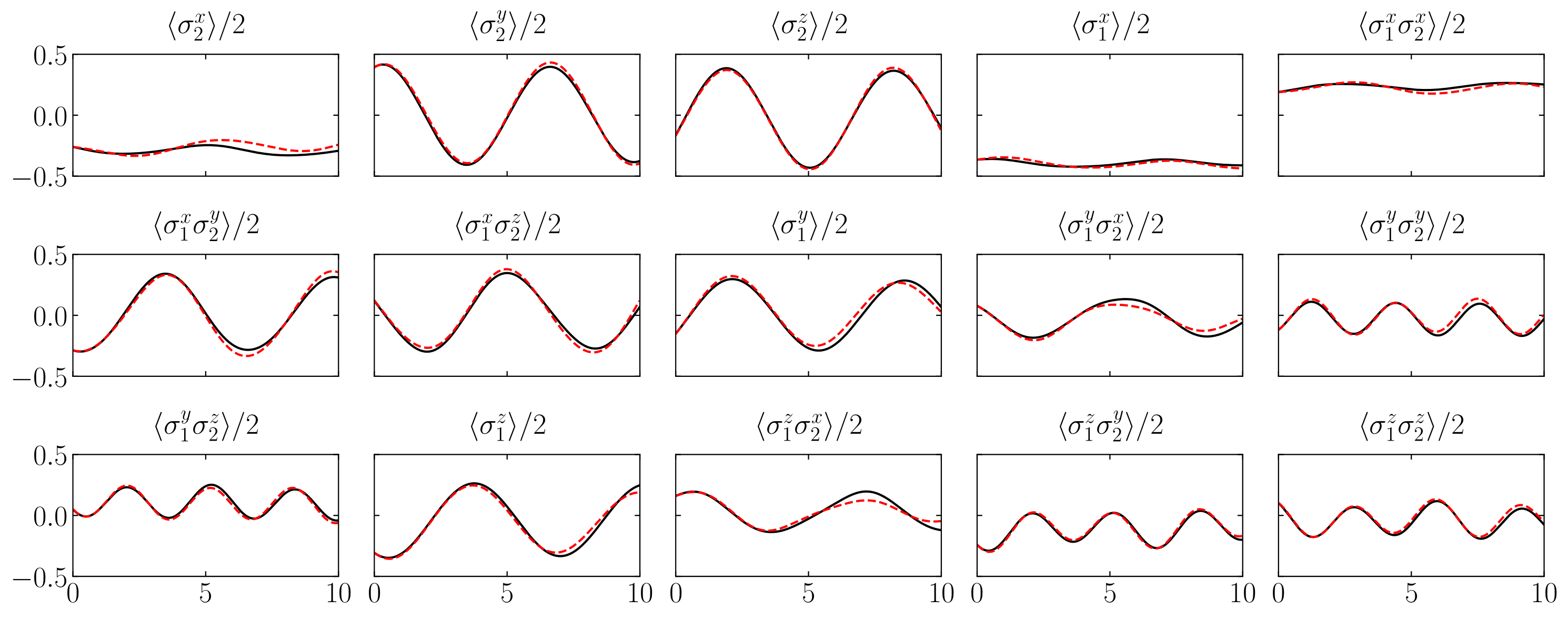}
    \caption{{\bf Full plot of the ML algorithm's performance.} 
    In figure the $15$ non-trivial components of the coherence vector $v_i$. The data refers to a spin chain with $V=0.1\Omega$, and $N=20, M=20$. The plots represent the exact time evolution (black solid line) and the dynamics predicted by the ML algorithm (red dashed line). In this case, the overall dynamics is roughly captured.}
    \label{fig_S_1}
\end{figure}
\begin{figure}[t!]
    \centering
    \includegraphics[scale=0.45]{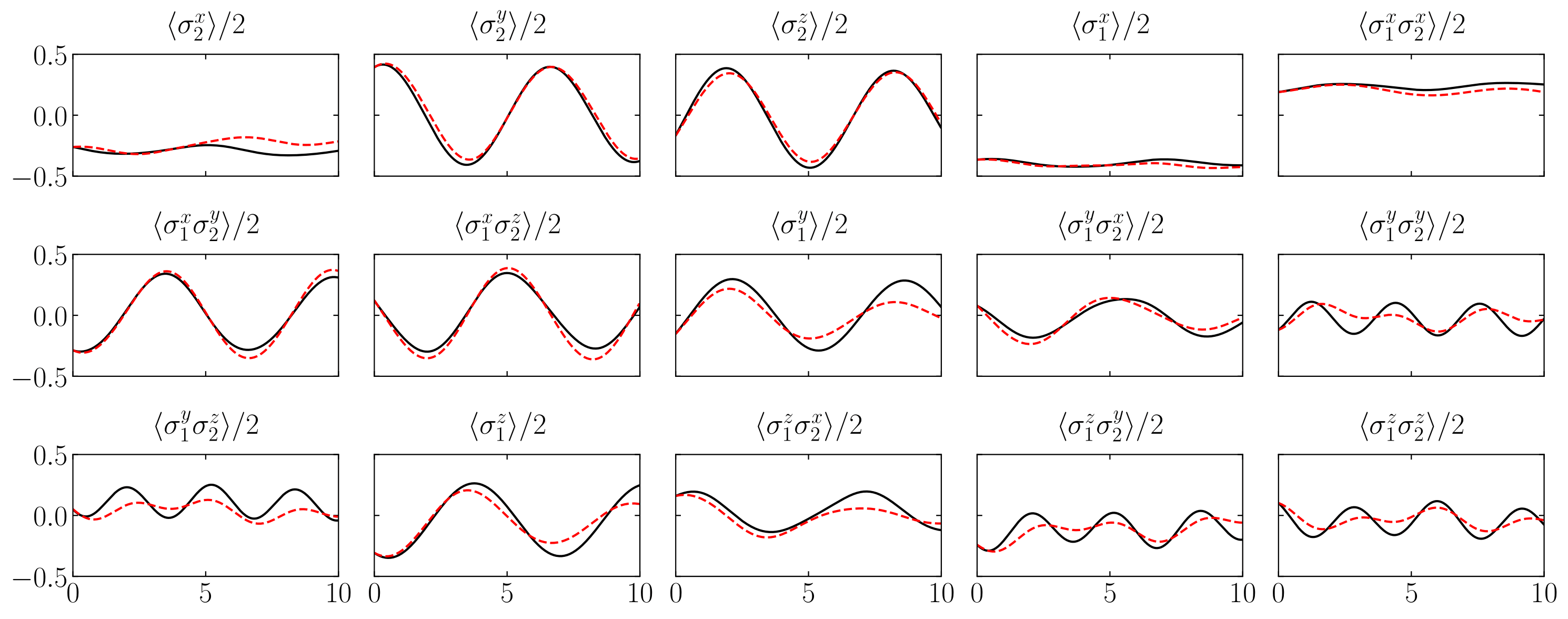}
    \caption{{\bf Full plot of the ML algorithm's performance.} 
    In figure the $15$ non-trivial components of the coherence vector $v_i$. The data refers to a spin chain with the same parameters as in Fig.~\ref{fig_S_1}, but $N=2, M=2$. The plots represent the exact time evolution (black solid line) and the dynamics predicted by the ML algorithm (red dashed line). In this case, the overall dynamics is well captured.}
    \label{fig_S_2}
\end{figure}
In Figure \ref{fig_S_1} and \ref{fig_S_2} we report the performance of the model trained on data that refers to a spin chain with $V=0.1\Omega$. In Fig. \ref{fig_S_1} the synthetic experimental data is generated with $N=20,M=20$. In this case, the ML prediction is almost perfectly overlapped on the exact line, indicating the correctness of the learned Lindbladian. In Fig. \ref{fig_S_2} the synthetic experimental data is generated with $N=2,M=2$. In this case, the dynamics is roughly captured, made exception for some observables e.g., $\sigma^y_1 \sigma^y_2, \sigma^y_1 \sigma^z_2, \sigma^z_1 \sigma^y_2 \text{ and } \sigma^z_1 \sigma^z_2,$.

In addition to the complete plots, we here report the learned expressions for the Hamiltonian and jump operators. In this case, we take into consideration the model trained over $N=20$ and $M=20$. For the Hamiltonian, we round up to two decimal places to improve readability; it reads
\begin{equation*}
    H = 0.5(\sigma^x_1 + \sigma^x_2) + 0.055(\sigma^z_1 + \sigma^z_2) + 0.025 \, \sigma^z_1 \sigma^z_2 - 0.03(\sigma^y_1 + \sigma^y_2) + 0.005( -\sigma^z_1 \sigma^x_2 - \sigma^y_1 \sigma^x_2 + \sigma^z_1 \sigma^y_2 - \sigma^x_1 \sigma^z_2 + \sigma^x_1 \sigma^y_2 ) \, .
\end{equation*}
Regarding dissipation part, there is only one non-zero eigenvalue (rounded to three decimal places) of the Kossakowski matrix $\gamma = 0.013$. Regarding the jump operator, rounded up to two decimal places, it reads
\begin{align*}
    J =  &0.245 \, \sigma^x_1 \sigma^z_2 + (0.19-0.1i) \, \sigma_2^y + (-0.16+0.07i) \, \sigma^x_1 \sigma^y_2 + (0.055+0.145i) \, \sigma_1^y \sigma_2^y + (0.095-0.075i) \, \sigma_1^y + \\
    & + (-0.09-0.08i) \, \sigma_1^y \sigma_2^x - (0.05+0.105i) \, \sigma_1^x + (0.09-0.065i) \, \sigma_1^x \sigma_2^x + (0.03-0.1i) \, \sigma^z_1 \sigma^x_2 + (-0.09+0.02i) \, \sigma_1^z + \\
    &+ (0.005+0.08i) \, \sigma_1^z \sigma_2^x + (-0.03+0.06i) \, \sigma_1^z \sigma_2^z +(0.02+0.035i) \, \sigma_1^y \sigma_2^z + 0.005 \, \sigma_2^z
\end{align*}

\begin{figure}[t!]
    \centering
    \includegraphics[scale=0.45]{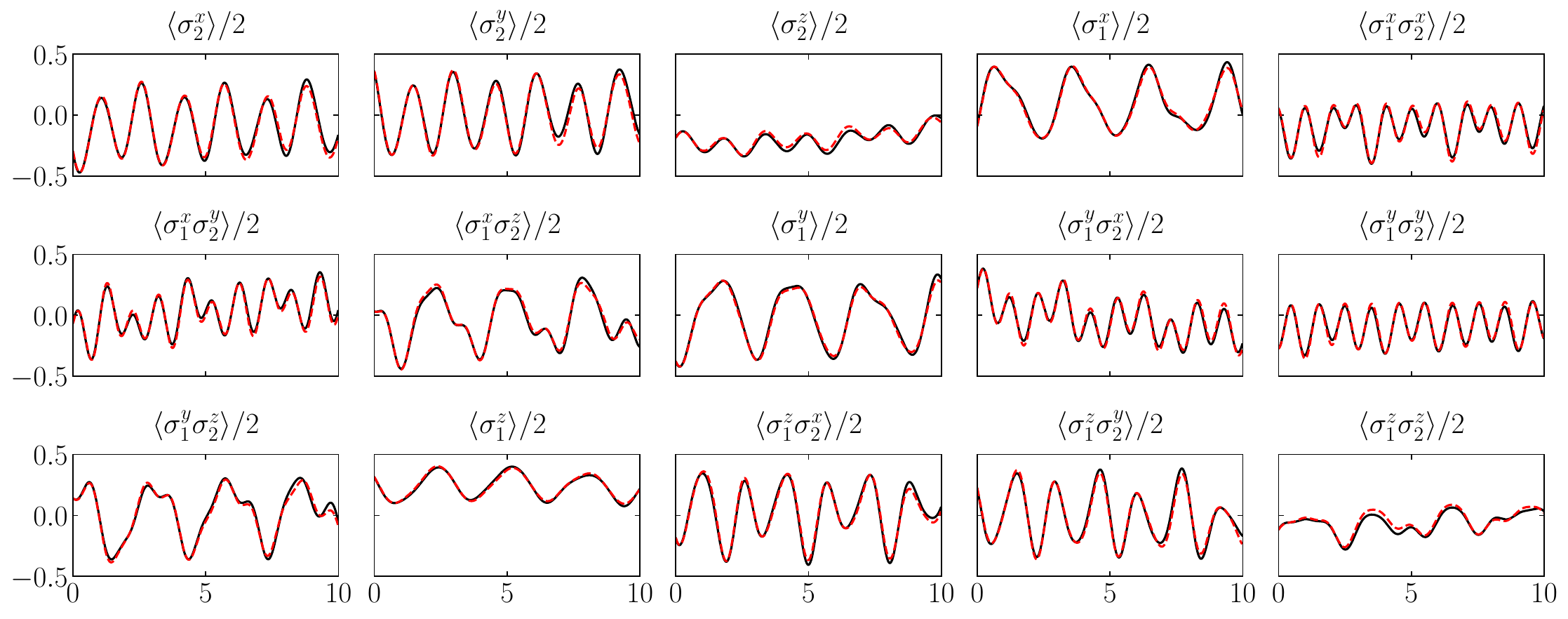}
    \caption{{\bf Full plot of the ML algorithm's performance.} 
    In figure the $15$ non-trivial components of the coherence vector $v_i$. The data refers to a spin chain with $V=2\Omega$, and $N=20, M=20$. The plots represent the exact time evolution (black solid line) and the dynamics predicted by the ML algorithm (red dashed line).In this case, the overall dynamics is well captured.}
    \label{fig_S_3}
\end{figure}
\begin{figure}[t!]
    \centering
    \includegraphics[scale=0.45]{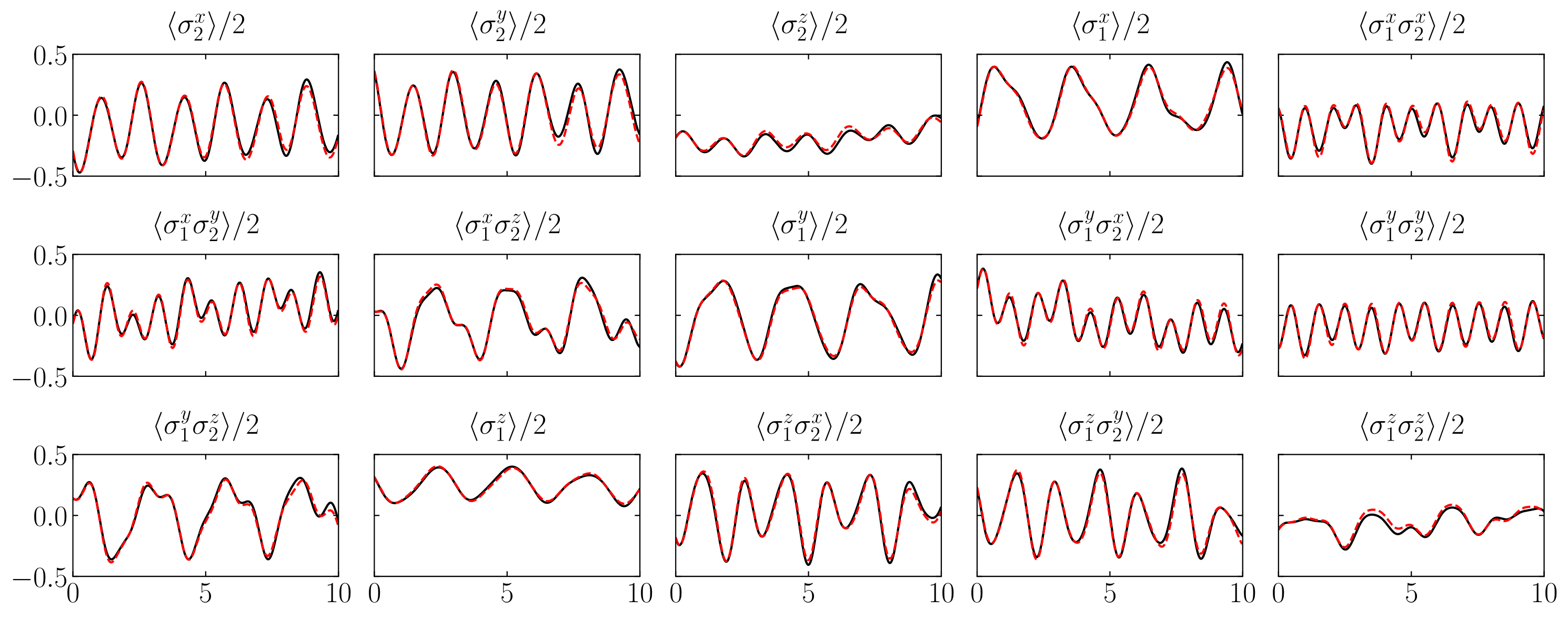}
    \caption{{\bf Full plot of the ML algorithm's performance.} 
    In figure the $15$ non-trivial components of the coherence vector $v_i$. The data refers to a spin chain with the same parameters as in Fig.~\ref{fig_S_3}, but $N=3, M=2$. The plots represent the exact time evolution (black solid line) and the dynamics predicted by the ML algorithm (red dashed line). In this case, the overall dynamics is porly captured.}
    \label{fig_S_4}
\end{figure}
In Figure \ref{fig_S_3} and \ref{fig_S_4} we report the performance of the model trained on data that refers to a spin chain with $V=2\Omega$. In Fig. \ref{fig_S_3} the synthetic experimental data is generated with $N=20,M=20$. In this case, the ML prediction is almost perfectly overlapped on the exact line, indicating the correctness of the learned Lindbladian. In Fig. \ref{fig_S_4} the synthetic experimental data is generated with $N=3,M=2$. In this case, the overall dynamics is poorly captured.

\end{document}